\documentclass[english]{article}
\usepackage[T1]{fontenc}
\usepackage[latin9]{inputenc}
\usepackage{geometry}
\usepackage{color}
\usepackage{babel}
\usepackage{amsmath}
\usepackage{amssymb}
\usepackage{graphicx}
\usepackage[unicode=true,pdfusetitle,
 bookmarks=true,bookmarksnumbered=false,bookmarksopen=false,
 breaklinks=false,pdfborder={0 0 0},pdfborderstyle={},backref=false,colorlinks=true]
 {hyperref}
\hypersetup{
 linkcolor=blue,citecolor=blue,urlcolor=blue}

\usepackage{cancel}

\newcommand{\bra}[1]{{\left\langle #1\right|}}
\newcommand{\skalarszorzat}[2]{{\langle #1 | #2 \rangle}}

\newcommand{\lanp}{{\boldsymbol u}^+_{N/2}}

\numberwithin{equation}{section}

\makeatletter
\usepackage{psfrag}

\makeatother

\begin{document}
\title{On factorized overlaps: Algebraic Bethe Ansatz, twists, and Separation
  of Variables}

\author{Tam\'as Gombor$^{1,2}$, Bal\'azs Pozsgay$^{1}$}

\date{\small $^1$\textit{Department of Theoretical Physics and \\
MTA-ELTE ``Momentum'' Integrable Quantum Dynamics Research Group\\
E\"otv\"os Lor\'and University \\
Budapest, Hungary}\\
  $^2$\textit{Wigner Research Centre for Physics\\ Budapest, Hungary}
  }

\maketitle

\begin{abstract}
We investigate the exact overlaps between eigenstates of integrable
spin chains and a special class of states called ``integrable
initial/final states''. These states satisfy a special integrability
constraint, and they are closely related to integrable boundary
conditions. 
We derive new algebraic relations for the integrable states, which lead
to a set of recursion relations for the exact overlaps. We solve
these recursion relations and thus we derive new overlap formulas,
valid in the XXX Heisenberg chain and its integrable higher spin
generalizations. Afterwards we generalize the integrability condition to twisted
boundary conditions, and derive the corresponding exact
overlaps. Finally, we embed the integrable states into the ``Separation
of Variables'' framework, and derive an alternative representation for
the exact overlaps of the XXX chain. Our derivations and proofs are
rigorous, and they can form the basis of future investigations
involving more complicated models such as nested or long-range
deformed systems.
\end{abstract}

\section{Introduction}

One dimensional quantum integrable models are very special systems, where the exact wave functions can be
computed using analytic methods, even in the presence of interactions. However, the computation of
composite objects such as correlation functions is a notoriously difficult task even in these
systems. In the last couple of years considerable effort was spent to study a special class of
objects: the exact overlaps between the eigenstates of the model and a distinguished set of states,
which are now called {\it integrable initial states}
\cite{sajat-integrable-quenches,sajat-mps}. These overlaps 
display very special features and they are important for a number of reasons.

First of all we mention the context of non-equilibrium dynamics of integrable
models, where the overlaps play a central role in the study of quantum quenches.
In a quench one
prepares the system in an initial state, which is not an eigenstate of the Hamiltonian, and the goal
is to study the non-equilibrium dynamics, the emergence of steady states, and their properties
\cite{essler-fagotti-quench-review}.  The overlaps with the initial state are intermediate objects
for the exact computations, and they serve as an input to the so-called Quench Action method, one of
the main methods for the study of the steady states \cite{quench-action}. Thus it was crucial to
find initial states where exact overlaps could be derived.
The first exact results appeared for the Lieb-Liniger model in 
\cite{caux-stb-LL-BEC-quench} and the Heisenberg spin chains in
\cite{sajat-neel,Caux-Neel-overlap1}. The exact overlap formulas played a central role in the early
studies of the Generalized Gibbs Ensemble (GGE) in interacting integrable models,
see \cite{JS-oTBA,sajat-oTBA}. 

It was found in these early studies that
the overlaps are non-zero
only if the eigenstate in question is an eigenvector of the space
reflection operator. Furthermore the overlaps were found to
possess a remarkable factorized form. It was later understood in \cite{sajat-integrable-quenches}
(see also \cite{sajat-minden-overlaps,sajat-mps}), that these properties are tied to special integrable
structures behind the initial states. Namely, the solvable cases can be understood as {\it integrable
  boundaries} in time, and they are closely related to integrable
boundary conditions.
This understanding was inspired by the seminal work of Ghoshal and
Zamolodchikov \cite{ghoshal-zamolodchikov}, which investigated
integrable boundaries in integrable Quantum Field Theories (QFT's).

In QFT the  overlaps between the
eigenstates and some finite volume integrable boundary state are called ``exact
$g$-functions''. The name ``$g$-function'' originally refers to
the overlap between  the finite volume ground state and the boundary
state, and this quantity has already a rather rich structure
\cite{Dorey:2004xk,sajat-g}. Overlaps with excited state are then
called excited state $g$-functions. We note that the factorized
structure of the excited state overlaps already appeared in
\cite{sajat-marci-boundary}, but this work did not influence the later
rigorous derivation of \cite{Caux-Neel-overlap1}.

The exact overlaps also appeared in the context of the AdS/CFT duality: it was first found in
\cite{zarembo-neel-cite1} that they describe one-point functions in defect CFT.
To be more precise, in a 
co-dimension one defect $\mathcal{N}=4$ SYM the tree-level one-point
functions are given by overlaps between
Bethe states corresponding to single trace
operators and special
two-site states
or Matrix Product States
\cite{zarembo-neel-cite3,ADSMPS2,kristjansen-proofs,adscft-kristjansen-2017,charlotte-zarembo-beyond-scalars}. On the string theory side the surface defect corresponds to
a probe brane, where closed strings corresponding to single trace
operators can be annihilated. The dual quantity of the one-point
function is the annihilation amplitude of the strings. In the 1+1
dimensional sigma model point of view this amplitude is an
overlap between and eigenstate and a boundary state, which is thus an excited
state $g$-function. For these reasons the overlaps appear on both the
gauge and the string theory sides of the duality.

For the probe
D5-brane, which preserves half of the supersymmetry it was showed that
the tree level boundary states are integrable in the $SO(6)$ sector,
and the corresponding overlaps were also calculated in
\cite{kristjansen-proofs}.
In \cite{adscft-kristjansen-2017} it was also showed that the one-point functions are integrable at one-loop level in the SU(2) sector.
Using symmetry considerations the
asymptotic all loop boundary state was proposed independently in
\cite{shota-valami} and \cite{gombor2020boundary1} and its asymptotic
overlaps was proposed for the full spectrum in
\cite{gombor2020boundary1,gombor2020boundary2}. In
\cite{charlotte-fermionic-dualities} the authors took the weak coupling limit of
the asymptotic overlap and showed that this formula is covariant under
the fermionic dualities and it is compatible with the tree level
overlaps in various subsectors
\cite{kristjansen-proofs,charlotte-zarembo-beyond-scalars}. This is a
strong consistency check of the assumption that the integrability
condition holds at the all loop level. 
It is worthwhile to mention that integrable boundary states also appear in
the investigation of three-point function where two operators are
of determinant type and third is of single trace type
\cite{Jiang:2019zig,yunfeng-structure-g}.  

Let us now discuss the various methods used
for the derivation of the exact overlaps.
The first work
\cite{caux-stb-LL-BEC-quench} applied coordinate Bethe Ansatz in the
Lieb-Liniger model to the
overlaps with low particle numbers, and a conjecture was made for the
general case. The paper \cite{Caux-Neel-overlap1} actually proved an exact
formula for the Heisenberg spin chains, valid for arbitrary particle
numbers. This proof was based on an off-shell overlap formula,
originally derived by Tsushiya \cite{tsushiya} and adapted to overlaps
in \cite{sajat-neel} (see also \cite{sajat-karol}). However, this
particular proof was valid only for a subclass of states, namely those
related to the so-called diagonal $K$-matrices. 
Afterwards a number of works simply just assumed
the factorized structure of the exact overlap, and determined the
pair amplitude of the overlaps using either coordinate Bethe Ansatz 
\cite{zarembo-neel-cite3,ADSMPS2,kristjansen-proofs,adscft-kristjansen-2017,charlotte-zarembo-beyond-scalars} or the so-called Quantum
Transfer Matrix (QTM) method
\cite{sajat-minden-overlaps,sajat-su3-2,sajat-twisted-yangian}.

A new approach was initiated in \cite{sajat-coba-proof}, where the
exact formulas were proven regarding all particle numbers, based on the
analytic properties of the coordinate Bethe Ansatz expressions. This
method is based on the ideas of Korepin, developed for the derivation
of the norm of the Bethe wave function \cite{korepin-norms}. It was
shown in \cite{sajat-coba-proof} that the method is applicable not
only to the Heisenberg chain, but also to spin chains with non-compact
local spaces such as the so-called $sl(2,\mathbb{R})$ chain. 
This method was already used in \cite{Chen:2020xel} to give an
alternative proof for  the
overlap formulas between the Lieb-Liniger Bethe states and the
Bose-Einstein condensate state, originally derived in
\cite{caux-stb-LL-BEC-quench} and first proven in \cite{Brockmann-BEC}.
However, the drawback of the method of \cite{sajat-coba-proof} is that it is relatively difficult to apply it
to new cases, and it was not clear whether it could be useful for the
models solvable by the nested Bethe Ansatz.

It is desirable to develop
further methods for the derivation and proof of exact overlap
formulas. It is fair to say that none of the existing methods is
convenient for actually proving the results in the nested spin
chains. Furthermore, it is desirable to find a formulation which would
allow an extension to long range deformed models, and in particular
to the full AdS/CFT problem.

With this goal in mind we turn to two central methods
used in integrability: the Algebraic Bethe Ansatz (ABA) and the Separation
of Variables (SoV). Remarkably, up to now neither of these methods have been
used for the systematic treatment of integrable initial states and their overlaps. The original work
\cite{sajat-neel} was based on some simple computations with systems
with boundaries in ABA, but the applicability of those results is very
limited. In integrable QFT the $g$-functions were considered in connection
with SoV in \cite{shota-g}, but that work did not treat the
integrability condition for the boundary states within SoV.
It is worthwhile to recall that the natural language of
the AdS/CFT spectral problem is the so-called $Q$-system, which is a set of relations for the
so-called $Q$-functions 
\cite{gromov-spectral-curve-intro}. On the other hand, the $Q$-functions are naturally interpreted
as the wave functions in SoV. This gives further motivation to apply the SoV method to the boundary states.
Nevertheless, a proper embedding of the integrable states into the SoV framework was missing up to
now.

In the present work we initiate a
systematic study of the overlaps using the ABA and the SoV
approaches. We show that the ABA is capable of deriving new recursion
relations for the overlaps, which can complement the methods of
\cite{sajat-coba-proof}, leading to a wider
applicability. Furthermore we show that the integrable initial states
can be represented also in the SoV framework, and we also derive new
overlap formulas using this approach.

\section{Overlaps and ABA}

In this section we review the construction of integrable two-site
states  for integrable spin chains \cite{sajat-integrable-quenches}. We focus on the XXX chain and
its integrable higher spin generalizations.

We introduce a new relation (which we call the KT-relation), which is very useful
for the derivation of overlap formulas: it leads to a new recursion
relation for the off-shell overlaps. This recursion can be solved in certain cases and we obtain the off-shell
overlap as a Vandermonde-like determinant. Using the argument of \cite{sajat-coba-proof} we derive the
on-shell formula of general two-site state for 
inhomogeneous chains with arbitrary quantum space. We close this section by calculating the overlaps
for descendant states. 

\subsection{Integrable states and KT relation}

Let us start with the definitions. Let us use the following convention
for the R-matrix
\begin{equation}
R(u)=u\mathbf{1}+\mathbf{P}=u\mathbf{1}+e_{ij}\otimes e_{ji},
\end{equation}
where $e_{ij}$-s are the $2\times2$ matrix units and $\mathbf{P}$
is the permutation matrix. Let $E_{ij}^{(r)}\in\mathrm{End}\left(\mathbb{C}^{r+1}\right)$
be the highest weight irrep of $\mathfrak{gl}_{2}$ with highest weight
$(r,0)$. The basis vectors are $\left|p,q\right\rangle ^{(r)}\in\mathbb{C}^{r+1}$
for which $p+q=r$, $0\leq p,q\leq r$ and
\begin{equation}
E_{11}^{(r)}\left|p,q\right\rangle ^{(r)}=p\left|p,q\right\rangle ^{(r)},\qquad E_{22}^{(r)}\left|p,q\right\rangle ^{(r)}=q\left|p,q\right\rangle ^{(r)}.
\end{equation}
Using these representations we define the Lax operators
\begin{equation}
R^{(r)}(u)=u\mathbf{1}+e_{ij}\otimes E_{ji}^{(r)},\qquad L^{(r)}(u)=R^{(r)}(u-r/2).
\end{equation}
The $R$- and $L$-matrices have crossing symmetry:
\begin{align}
R_{12}(u) & =-\sigma_{1}^{y}R_{12}^{t_{1}}(-u-1)\sigma_{1}^{y}.\\
L_{12}^{(r)}(u) & =-\sigma_{1}^{y}\left[L_{12}^{(r)}(-u)\right]^{t_{1}}\sigma_{1}^{y}=-\Sigma_{2}\left[L_{12}^{(r)}(-u)\right]^{t_{2}}\Sigma_{2}^{-1},
\end{align}
where $\sigma^{y}=\left(\begin{array}{cc}
0 & -i\\
i & 0
\end{array}\right)$ and the $\Sigma$ is a matrix for which
\begin{align}
\Sigma\left(E_{11}^{(r)}-E_{22}^{(r)}\right)\Sigma^{-1} & =E_{22}^{(r)}-E_{11}^{(r)}, & \Sigma E_{12}^{(r)}\Sigma^{-1} & =-E_{21}^{(r)}, & \Sigma E_{21}^{(r)}\Sigma^{-1} & =-E_{12}^{(r)}.
\end{align}
Using the Lax operators we can define the monodromy and transfer matrices
and their space reflected versions as
\begin{equation}
  \label{Tdef}
  \begin{split}
    T_{0}(u)&=L_{0,2L}^{(r_{2L})}(u-\xi_{2L})L_{0,2L-1}^{(r_{2L-1})}(u-\xi_{2L-1})\dots
    L_{0,2}^{(r_{2})}(u-\xi_{2})L_{0,1}^{(r_{1})}(u-\xi_{1})\\
T_{0}^{\pi}(u)&=    L_{0,1}^{(r_{1})}(u+\xi_{1})L_{0,2}^{(r_{2})}(u+\xi_{2})\dots L_{0,2L-1}^{(r_{2L-1})}(u+\xi_{2L-1})L_{0,2L}^{(r_{2L})}(u+\xi_{2L})\\
  \end{split}
\end{equation}
and
\begin{align}
t(u)  =\mathrm{Tr}_{0}T_{0}(u),\qquad
\Pi t(u)\Pi  =\mathrm{Tr}_{0}T_{0}^{\pi}(u),
\end{align}
where $\Pi$ is the space reflection operator. The matrix entries
of the monodromy matrices are
\begin{align}
T(u) & =\left(\begin{array}{cc}
A(u) & B(u)\\
C(u) & D(u)
\end{array}\right), & T^{\pi}(u) & =\left(\begin{array}{cc}
A^{\pi}(u) & B^{\pi}(u)\\
C^{\pi}(u) & D^{\pi}(u)
\end{array}\right).
\end{align}
The $T$ and $T^{\pi}$ are not independent quantities.
We can use the crossing symmetry of Lax matrices to express
\begin{equation}
\sigma_{0}^{y}\left(T_{0}^{\pi}(u)\right)^{t_{0}}\sigma_{0}^{y}=L_{0,2L}(-u-\xi_{2L})\dots L_{0,1}(-u-\xi_{1})=T_{0}(-u).
\end{equation}
Therefore
\begin{equation}
T^{\pi}(u)=\left(\begin{array}{cc}
D(-u) & -B(-u)\\
-C(-u) & A(-u)
\end{array}\right).\label{eq:crossMon}
\end{equation}
We can construct the eigenvalues and eigenvectors of 
the transfer matrix using the algebraic Bethe Ansatz. The eigenvectors can be built form the $B$-operators as
\begin{equation}
\left|\mathbf{u}\right\rangle =B(u_{1})\dots B(u_{N})\left|0\right\rangle,
\end{equation}
where $\left|0\right\rangle $ is the reference state:
\begin{equation}
\left|0\right\rangle =\left|r_{1},0\right\rangle ^{(r_{1})}\otimes\left|r_{2},0\right\rangle ^{(r_{2})}
\otimes\dots\otimes
\left|r_{2L-1},0\right\rangle ^{(r_{2L-1})}\otimes\left|r_{2L},0\right\rangle ^{(r_{2L})}.
\end{equation}
Below it is understood that $\mathbf{u}$ is a set of rapidities with $N$ elements, unless
otherwise noted.

The action of operators $A(u)$ and $D(u)$ on the pseudovacuum can be written in the usual form
\begin{equation}
A(u)\left|0\right\rangle   =\lambda^{+}(u)\left|0\right\rangle , \qquad
D(u)\left|0\right\rangle   =\lambda^{-}(u)\left|0\right\rangle,
\end{equation}
where
\begin{equation}
  \label{lambdapm}
 \lambda^{\pm}(u) = \prod_{k=1}^{2L}\left(u \pm r_{k}/2-\xi_{k}\right).
\end{equation}
The vector $\left|\mathbf{u}\right\rangle$ is an eigenvector of the transfer matrix is the Bethe roots $u_i$ satisfy the Bethe Ansatz equations
\begin{equation}
 \frac{\lambda^+(u_i)}{\lambda^-(u_i)} = - \frac{Q_1(u_i+1)}{Q_1(u_i-1)},
\end{equation}
and the eigenvalue $\Lambda(u)$ reads as
\begin{equation}
  \Lambda(u)=\lambda^{+}(u)\frac{Q_{1}(u-1)}{Q_{1}(u)}+\lambda^{-}(u)\frac{Q_{1}(u+1)}{Q_{1}(u)},
  \label{eq:Lam}
\end{equation}
where we defined the $Q$-function
\begin{equation}
Q_{1}(u)=\prod_{i=1}^{N}(u-u_{i}).
\end{equation}

We can also construct the left eigenvectors with the same eigenvalue as
\begin{equation}
\left\langle \mathbf{u}\right| = \left\langle 0 \right| C(u_{1})\dots C(u_{N}).
\end{equation}
In this paper we do not use any inner product therefore the quantities
like $\langle \mathbf{v}|\mathbf{u}\rangle$ are defined as a natural
pairing between a dual vector and a vector, but not as a scalar
product. 

It is well known that the Bethe states with $N$ magnons are highest
weight states and their descendants are also eigenvectors of the transfer
matrix with the same eigenvalue. Let us define spin operators as
\begin{equation}
S^{+}=e_{21},\qquad S^{-}=e_{12},\qquad S_{3}=\frac{1}{2}\left(e_{11}-e_{22}\right),
\end{equation}
for which
\begin{equation}
\left[S^{+},S^{-}\right]=2S_{3},\qquad\left[S_{3},S^{\pm}\right]=\pm S^{\pm}.
\end{equation}
The asymptotic limit of the operator $B$ is the spin lowering operator
\begin{equation}
\lim_{u\to\infty}\frac{1}{u^{2L-1}}B(u)=\Delta(S^{-}),
\end{equation}
where $\Delta$ is the usual co-product. Let us define the descendant
states as
\begin{equation}
\left|\mathbf{u},M\right\rangle :=\Delta(S^{-})^{M}\left|\mathbf{u}\right\rangle.
\end{equation}

Below we will construct integrable two-site states. The construction requires that 
the sites of the spin chain are ``paired'' which means that
for a pair of sites $(2a-1,2a)$ the inhomogeneities are opposite
to each other i.e.
\begin{equation}
\xi_{2a-1}=\theta_{a},\qquad\xi_{2a}=-\theta_{a}
\end{equation}
and the representations are the same i.e.
\begin{equation}
r_{2a-1}=r_{2a}=s_{a}.
\end{equation}
Two-site states can be built from $K$-matrices $K^{(s)}(u)$. The $K$-matrices are given by their
components $k_{ij}$, which can be arranged into a matrix or into a co-vector as
\begin{equation}
\mathbf{K}_{1}(u)=\sum_{i,j}k_{ij}(u)e_{ij},\qquad \vec{K}_{12}(u)=\sum_{i,j}k_{ij}(u)e_{i}^{*}\otimes e_{j}^{*}.
\end{equation}
These are two representations of the same quantity and we for simplicity we call them both
$K$-matrix.

\begin{figure}
\begin{centering}
\includegraphics[width=0.8\textwidth]{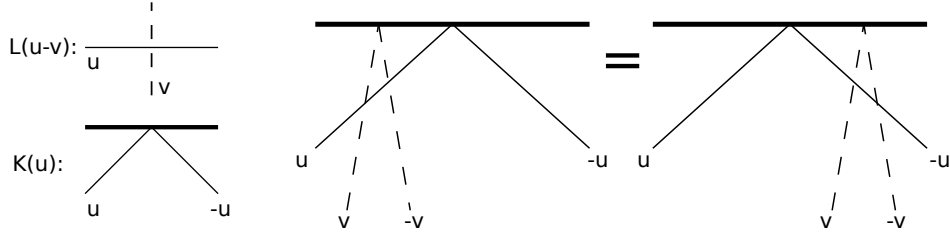}
\par\end{centering}
\caption{Pictorial representation of the KYB equation \eqref{eq:KYB}, see on the right. The action of the Lax operator is represented
by the crossing of the straight and dashed lines, which correspond to two (potentially different)
auxiliary spaces. The thick horizontal line represents for the integrable final state (boundary in
the time direction). The $K$-matrix connects the physical spaces to the integrable boundary, and its
two ``legs'' carry rapidity parameters (inhomogeneities) opposite to each other.}

\label{fig:KYB}
\end{figure}

The defining equation of the $K$-matrices is the following
equation \cite{sajat-integrable-quenches,sajat-mps,gombor2020boundary1,gombor2020boundary2}
\begin{equation}
\vec{K}_{12}(u)\vec{K}_{34}^{(s)}(v)L_{14}^{(s)}(u+v)L_{13}^{(s)}(u-v)=\vec{K}_{12}(u)\vec{K}_{34}^{(s)}(v)L_{23}^{(s)}(u+v)L_{24}^{(s)}(u-v).\label{eq:KYB}
\end{equation}
Following \cite{gombor2020boundary1,gombor2020boundary2} we call it
the KYB equation. We will see below that it is very closely related to
the Boundary Yang-Baxter (BYB) equation. A graphical interpretation of
the KYB equation is given in Figure \ref{fig:KYB}.

We can now define two-site states as 
\begin{equation}
\left\langle \Psi\right|=\vec{K}^{(s_{1})}(\theta_{1})\otimes\dots \otimes \vec{K}^{(s_{L})}(\theta_{L}).
\end{equation}
We call such a state an {\it integrable final state}. We use the expression
 ``final state'', because in our conventions it is actually a co-vector
which acts on the eigenstates of the model.

\begin{figure}
\begin{centering}
\includegraphics[width=0.9\textwidth]{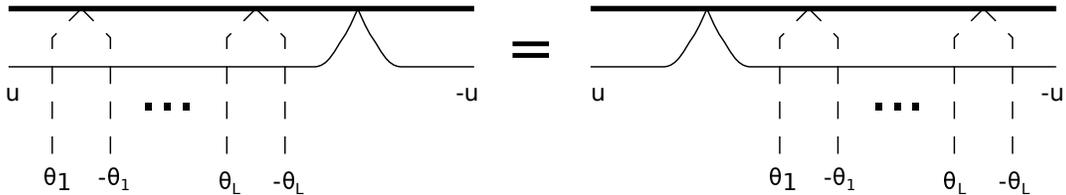}
\par\end{centering}
\caption{Graphical interpretation of the KT relation \eqref{eq:KT}. The dashed vertical lines stand
  for the physical spaces of the spin chain, and the horizontal line is the auxiliary space of the
  monodromy matrix. We assume that this space carries the spin-1/2 representation, but the physical
  spaces can be higher dimensional. The main idea is to insert one more $K$-matrix which acts on
  the auxiliary space, and then to use the KYB relations consecutively, to shift the action of the
  extra $K$ from the right hand side to the left hand side. Working out the details we obtain
  \eqref{eq:KT}. It is important that the indices at the left and right are free, thus we actually
  have a collection of 4 algebraic relations. The integrability condition \eqref{eq:intCond} is
  found by multiplying with the inverse of one of the $K$-matrices, and taking the trace in
  auxiliary space.
  }

\label{fig:KT}
\end{figure}

It follows directly from the KYB equation that such a two-site state satisfies the
following relation (see \ref{fig:KT} for the graphical derivation)
\begin{equation}
\left\langle \Psi\right|\vec{K}_{12}(u)T_{1}(u)=\left\langle \Psi\right|\vec{K}_{12}(u)T_{2}^{\pi}(u) \quad \Longleftrightarrow \quad
\left\langle \Psi\right|\sigma^{y}\mathbf{K}^{t}(u)T(u)=
\left\langle  \Psi\right|T(-u)\sigma^{y}\mathbf{K}^{t}(u).
\label{eq:KT}
\end{equation}
We call \eqref{eq:KT} the $KT$-relation; it can be understood as the ``time''-boundary analog of the
usual $RTT$-relation. Note that it is actually a collection of 4 equations due to the free indices in
the auxiliary space; specific components will be given below. Also, it is shown below that the
$KT$-relation allows us to 
embed the overlap computations into the 
framework of algebraic Bethe Ansatz.

Let us assume that the $K$-matrix is invertible. In this case we can take the trace of the KT
relation in auxiliary space after multiplying with the inverse of the $K$-matrix on either side. This
leads to the integrability
condition 
\begin{equation}
  \left\langle \Psi\right|t(u)=\left\langle \Psi\right|\Pi t(u)\Pi.
  \label{eq:intCond}
\end{equation}
This condition was introduced in \cite{sajat-integrable-quenches},
and the equation \eqref{eq:KT} appears there as an intermediate step in the derivation
\eqref{eq:intCond}.
In this paper we point out that the equation \eqref{eq:KT} contains more information than the
integrability condition \eqref{eq:intCond}, and we use this extra information
in the derivation of the overlaps.
We note that 
\eqref{eq:intCond} is formally a relation for co-vectors.

It is known that the KYB equation is
equivalent to the reflection equation \cite{sajat-integrable-quenches,sajat-mps,gombor2020boundary1,gombor2020boundary2}
\begin{equation}
  L_{12}^{(s)}(u-v)\left(\Sigma_{1}\mathbf{K}_{1}^{(s)}(-u)\right)L_{12}^{(s)}(u+v)\left(\sigma_{2}^{y}\mathbf{K}_{2}(-v)\right)=\left(\sigma_{2}^{y}\mathbf{K}_{2}(-v)\right)L_{12}^{(s)}(u+v)\left(\Sigma_{1}\mathbf{K}_{1}^{(s)}(-u)\right)L_{12}^{(s)}(u-v).
  \label{eq:BYBE}
\end{equation}
For $s=1$ the reflection equation can be written as
\begin{multline}
R_{12}(u-v-1/2)\left(\sigma_{1}^{y}\mathbf{K}_{1}^{(1)}(-u)\right)R_{12}(u+v-1/2)\left(\sigma_{2}^{y}\mathbf{K}_{2}(-v)\right)=\\
\left(\sigma_{2}^{y}\mathbf{K}_{2}(-v)\right)R_{12}(u+v-1/2)\left(\sigma_{1}^{y}\mathbf{K}_{1}^{(1)}(-u)\right)R_{12}(u-v-1/2).
\end{multline}
Using the definitions
\begin{equation}
\vec{K}^{(1)}(u):=\frac{1}{u+1/2}\vec{K}(u+1/2),\qquad\mathcal{K}_{1}(u):=\sigma_{1}^{y}\mathbf{K}_{1}(-u)
\end{equation}
we obtain the usual reflection equation of XXX spin chain
\begin{equation}
R_{12}(u-v)\mathcal{K}_{1}(u)R_{12}(u+v)\mathcal{K}_{2}(v)=\mathcal{K}_{2}(v)R_{12}(u+v)\mathcal{K}_{1}(u)R_{12}(u-v).
\end{equation}
The solutions of this equation are well known. We use the following
parametrization
\begin{align}
k_{11}(u) & =au & k_{12}(u) & =-b+cu \label{eq:genK1}\\
k_{21}(u) & =b+cu & k_{22}(u) & =du. \label{eq:genK2}
\end{align}
We will also use normalized matrix entries
\begin{equation}
\kappa(u):=\frac{k_{21}(u)}{k_{11}(u)}=\frac{b+cu}{au},\quad\frac{k_{12}(u)}{k_{11}(u)}=\frac{-b+cu}{au}=\kappa(-u),\quad\delta:=\frac{k_{22}(u)}{k_{11}(u)}=\frac{d}{a}.
\end{equation}

\begin{figure}
\begin{centering}
\includegraphics[width=0.4\textwidth]{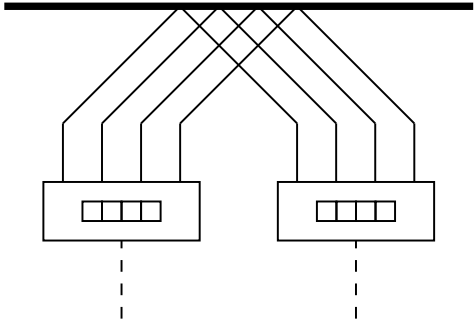}
\par\end{centering}
\caption{Graphical illustration of the fusion rule \eqref{eq:Kfusion} for the $K$-matrices. Here we present the example
  $K^{(4)}(v)$. The fused $K$-matrix is obtained by taking 4 copies of the original $K$-matrices
  with shifted rapidity parameters. Afterwards we apply a projection, which is symbolized
  by the boxes.}
\label{fig:Kfusion}
\end{figure}

Using this matrix we can obtain the $K$-matrices for the higher representations
by the fusion procedure. It is more convenient to use the reflection
equation (\ref{eq:BYBE}) with the $R$-matrices. We use the $s=1$
equation
\begin{equation}
\vec{K}_{12}^{(1)}(u)\vec{K}_{34}^{(1)}(v)R_{14}(u+v)R_{13}(u-v)=\vec{K}_{12}^{(1)}(u)\vec{K}_{34}^{(1)}(v)R_{23}(u+v)R_{24}(u-v)
\end{equation}
to construct the solution of the general KYB equation
\begin{equation}
\vec{K}_{12}^{(1)}(u)\vec{K}_{34}^{(s)}(v)L_{14}^{(s)}(u+v+1/2)L_{13}^{(s)}(u-v+1/2)=\vec{K}_{12}^{(1)}(u)\vec{K}_{34}^{(s)}(v)L_{23}^{(s)}(u+v+1/2)L_{24}^{(s)}(u-v+1/2).
\end{equation}
The $L_{14}^{(s)}(u\mp v+1/2)$ can be obtained from the original $R$-matrix
using fusion \cite{resh-fusion}. We obtain 
\begin{equation}
L_{ab}^{(s)}(u\mp v+1/2)\propto\left[\mathcal{P}_{b}^{(1,2,\dots,s)}\right]\left[\prod_{k=1}^{s}R_{a,k}(u\mp(v-\frac{s-2k+1}{2}))\right]\left[\mathcal{P}_{b}^{(1,2,\dots,s)}\right]^{T},
\end{equation}
where $\mathcal{P}^{(1,\dots,s)}:\left(\mathbb{C}^{2}\right)^{\otimes s}\to\mathbb{C}^{s+1}$
is the projection operator to the symmetric subspace of $\left(\mathbb{C}^{2}\right)^{\otimes s}$.
The $K^{(s)}(v)$ can be obtained in an analogous way  \cite{fusion-open-chains}
\begin{multline}
\vec{K}_{a,b}^{(s)}(v)=\left[\prod_{k=1}^{s}\vec{K}_{2k-1,2k}^{(1)}(v-\frac{s-2k+1}{2})\right]\times\\
\left[\prod_{l=1}^{s-1}\prod_{k=1}^{s-l}\bar{R}_{2k+2l-1,2k}(2v-s+2k+l-1)\right]\left[\mathcal{P}_{a}^{(1,3,\dots,2s-1)}\right]^{T}\left[\mathcal{P}_{b}^{(2,4,\dots,2s)}\right]^{T},\label{eq:Kfusion}
\end{multline}
where
\begin{equation}
\bar{R}(u)=\frac{1}{u+1}R(u).
\end{equation}
In figure \ref{fig:Kfusion} we present a graphical interpretation of the fusion, taking the example
of $\vec{K}_{a,b}^{(4)}(v)$. 

 Using the this convention for the $K$-matrices the overlap between $\bigl\langle\Psi\bigr|$ and the pseudovacuum is
\begin{equation}
\bigl\langle\Psi\bigr|0\bigr\rangle=a^{\mathcal{L}},
\end{equation}
where
\begin{equation}
\mathcal{L}=\sum_{k=1}^{L}s_{k}.
\end{equation}

In this section our goal is finding and proving overlap formulas as 
\begin{equation}
 \bigl\langle\Psi\bigr|\mathbf{u}\bigr\rangle, \label{eq:ov}
\end{equation}
where $\bigr|{\bf u}\bigr\rangle$ is an on-shell vector i.e. an
eigenstate of the transfer matrices. However, quantities like
\eqref{eq:ov} are not physical since they depend on our normalization
conventions. Of course, for physical system the Hamiltonian is
hermitian and we have a well-defined norm on the Hilbert space and we
can divide \eqref{eq:ov} by the norm of the state. In this paper we do
not introduce any metric but we have an option to define normalized
quantities as  
\begin{equation}
 \frac{\bigl\langle\Psi\bigr|\mathbf{u}\bigr\rangle \bigl\langle\mathbf{u}\bigr|\Psi^*\bigr\rangle}{\bigl\langle\mathbf{u}\bigr|\mathbf{u}\bigr\rangle},
\end{equation}
where $|\mathbf{u}\rangle$ and $\langle\mathbf{u}|$ are the right and
left eigenvectors with the same eigenvalue (and spin). The advantage
of this expression is that it does not depend on the normalization of
the left/right eigenvectors. The disadvantage is that we have to define a
new initial two-site state $\bigr|\Psi^*\bigr\rangle$ which is
independent from the final two-site state $\bigl\langle\Psi\bigr|$
(because of the absence of the metric). With the asterisk we suggest that
it would be a complex conjugated state in a physical situation. 

In the general case we can also define the initial two-site states. We can use
the vector rearrangement of the previously defined $K$-matrices 
\begin{equation}
 \vec{K}_{12}(u)^T=\sum_{i,j}k_{ij}(u)e_{i}\otimes e_{j}.
\end{equation}
We saw that they satisfy the reflection equation \eqref{eq:BYBE}
\begin{equation}
L_{12}^{(s)}(u-v)\left(\Sigma_{1}\mathbf{K}_{1}^{(s)}(-u)\right)L_{12}^{(s)}(u+v)\left(\sigma_{2}^{y}\mathbf{K}_{2}(-v)\right) = 
\left(\sigma_{2}^{y}\mathbf{K}_{2}(-v)\right)L_{12}^{(s)}(u+v)\left(\Sigma_{1}\mathbf{K}_{1}^{(s)}(-u)\right)L_{12}^{(s)}(u-v).
\end{equation}
This equation can be written as
\begin{equation}
L_{13}^{(s)}(v-u) L_{14}^{(s)}(-u-v) 
 \vec{K}_{12}(-u)^T 
 \vec{K}_{34}^{(s)}(-v)^T  = 
L_{24}^{(s)}(v-u) L_{23}^{(s)}(-u-v) 
 \vec{K}_{12}(-u)^T 
 \vec{K}_{34}^{(s)}(-v)^T,
\end{equation}
which will be called KYB equation for right vectors.
The initial two-site state is
\begin{equation}
\bigr|\Psi \bigr\rangle =
\vec{K}^{(s_{1})}(-\theta_{1})^T
\otimes \dots \otimes 
\vec{K}^{(s_{L})}(-\theta_{L})^T,
\end{equation}
for which the corresponding $KT$-relation can be written as
\begin{equation}
T_{2}(u) \vec{K}_{12}(u)^T \left| \Psi\right\rangle=
T_{1}^{\pi}(u) \vec{K}_{12}(u)^T \left| \Psi\right\rangle \quad \Longleftrightarrow \quad
T(u) \mathbf{K}^{*t}(u) \sigma^{y} \left| \Psi\right\rangle
= \mathbf{K}^{*t}(u) \sigma^{y} T(-u)\left| \Psi\right\rangle.
\end{equation}
In the following we concentrate overlaps with the final two site
states but we emphasize that the calculation for the initial two-site
state is completely analogous. 

\subsection{The $KT$ relation and the overlaps}

In the following we will use the $KT$ relation to calculate overlaps
between the above defined two-site states $\left\langle \Psi\right|$
and the off-shell Bethe states.
The main advantage of this method is that it can be applied even for
inhomogeneous chains and it gives the overlap formula for all representations
of the quantum space simultaneously. This is remarkable, because in the higher spin cases the
$K$-matrices are not known explicitly, they are given by fusion \eqref{eq:Kfusion}. Nevertheless it is
possible to derive the exact overlaps.

We start with the following component of the KT relations:
\begin{equation}
\left\langle
  \Psi\right|\left(k_{11}(u)B(u)+k_{21}(u)D(u)\right)=\left\langle
  \Psi\right|\left(k_{21}(u)A^{\pi}(u)+k_{22}(u)C^{\pi}(u)\right). 
\end{equation}
Using the crossing relation for the space reflected monodromy matrix
(\ref{eq:crossMon}) we obtain 
\begin{equation}
\left\langle \Psi\right|\left(k_{11}(u)B(u)+k_{21}(u)D(u)\right)=\left\langle \Psi\right|\left(k_{21}(u)D(-u)-k_{22}(u)C(-u)\right).\label{eq:rem}
\end{equation}

We can see that this equation can be used to replace $B$-operators
with $D$- and $C$-operators, therefore we can decrease the number of magnons iteratively.
Using the normalized matrix entries $\kappa(u)$ and $\delta$ the
equation (\ref{eq:rem}) can be written as
\begin{equation}
  \label{BDC}
\left\langle \Psi\right|B(u)=\kappa(u)\left\langle \Psi\right|\left(D(-u)-D(u)\right)-\delta\left\langle \Psi\right|C(-u).
\end{equation}
This will lead to a recursion relation for the overlaps, as we show
below. We note that a different type of recursion was earlier used in
\cite{calabrese-recursive-overlaps}, where the recursion also involved
changing the length of the spin chain in each step. In contrast, the
relation \eqref{BDC} does not change the volume, thus it leads to a
different recursion procedure.

Now we demonstrate the usefulness of the KT-relation
by very simple and instructive calculations which give the off-shell overlaps for one and two magnons. 

\subsection*{N=1}

We calculate the one magnon overlap as
\begin{equation}
\left\langle \Psi\right|B(u)\left|0\right\rangle =\kappa(u)\left\langle \Psi\right|\left(D(-u)-D(u)\right)\left|0\right\rangle -A\left\langle \Psi\right|C(-u)\left|0\right\rangle =a^{\mathcal{L}}\kappa(u)\left(\lambda^{+}(u)-\lambda^{-}(u)\right),
\end{equation}
where we used the identities
\begin{equation}
\lambda^{\pm}(-u)=\lambda^{\mp}(u).
\end{equation}

\subsection*{N=2}

Let us calculate the two magnon overlap
\begin{equation}
\left\langle \Psi\right|B(u_{1})B(u_{2})\left|0\right\rangle =\kappa(u_{1})\left\langle \Psi\right|\left(D(-u_{1})-D(u_{1})\right)B(u_{2})\left|0\right\rangle -\delta\left\langle \Psi\right|C(-u_{1})B(u_{2})\left|0\right\rangle .
\end{equation}
Using the commutation relations
\begin{align}
D(u)B(v) & =\frac{u-v+1}{u-v}B(v)D(u)-\frac{1}{u-v}B(u)D(v)\\
C(u)B(v) & =B(v)C(u)+\frac{1}{u-v}\left(A(v)D(u)-A(u)D(v)\right)
\end{align}
we obtain the following expressions
\begin{multline}
\left\langle \Psi\right|D(u_{1})B(u_{2})\left|0\right\rangle =\frac{u_{1}-u_{2}+1}{u_{1}-u_{2}}\lambda^{-}(u_{1})\left\langle \Psi\right|B(u_{2})\left|0\right\rangle -\frac{1}{u_{1}-u_{2}}\lambda^{-}(u_{2})\left\langle \Psi\right|B(u_{1})\left|0\right\rangle =\\
a^{\mathcal{L}}\Biggl[\kappa(u_{2})\frac{u_{1}-u_{2}+1}{u_{1}-u_{2}}\lambda^{-}(u_{1})\left(\lambda^{+}(u_{2})-\lambda^{-}(u_{2})\right)-\kappa(u_{1})\frac{1}{u_{1}-u_{2}}\lambda^{-}(u_{2})\left(\lambda^{+}(u_{1})-\lambda^{-}(u_{1})\right)\Biggr]
\end{multline}
and
\begin{equation}
\left\langle \Psi\right|C(-u_{1})B(u_{2})\left|0\right\rangle =-a^{\mathcal{L}}\frac{1}{u_{1}+u_{2}}\left(\lambda^{+}(u_{1})\lambda^{+}(u_{2})-\lambda^{-}(u_{1})\lambda^{-}(u_{2})\right).
\end{equation}
Therefore the two magnon overlap can be written as
\begin{align}
\left\langle \Psi\right|B(u_{1})B(u_{2})\left|0\right\rangle =a^{\mathcal{L}}\Biggl[ & \left(\kappa(u_{1})\kappa(u_{2})\frac{u_{1}+u_{2}-1}{u_{1}+u_{2}}+\delta\frac{1}{u_{1}+u_{2}}\right)\lambda^{+}(u_{1})\lambda^{+}(u_{2})-\nonumber \\
- & \kappa(u_{1})\kappa(u_{2})\frac{u_{1}-u_{2}+1}{u_{1}-u_{2}}\lambda^{-}(u_{1})\lambda^{+}(u_{2})-\nonumber \\
- & \kappa(u_{1})\kappa(u_{2})\frac{u_{1}-u_{2}-1}{u_{1}-u_{2}}\lambda^{+}(u_{1})\lambda^{-}(u_{2})+\nonumber \\
+ & \left(\kappa(u_{1})\kappa(u_{2})\frac{u_{1}+u_{2}+1}{u_{1}+u_{2}}-\delta\frac{1}{u_{1}+u_{2}}\right)\lambda^{-}(u_{1})\lambda^{-}(u_{2})\Biggr].
\end{align}

\subsection{Global rotation}

The previous examples showed that the term in the $KT$-relation proportional to $\delta$ causes
significant difficulty. This arises from the appearance of the $C$-operators, whose commutation
relations with the $B$-operators are more involved.

However, we can use a global rotation to eliminate the $\delta$-terms. Applying the following rotation
\begin{equation}
  \label{rotK}
 \vec{K}'_{12}(u) = \vec{K}_{12}(u) \exp\left(\frac{\gamma-c}{a} S^+\right),
\end{equation}
where 
\begin{equation}
 \gamma^2 = c^2 - ad
\end{equation}
the components of the rotated $K$ matrix are 
\begin{align}
k'_{11}(u) & =au & k'_{12}(u) & =-b + \gamma u\\
k'_{21}(u) & =b+\gamma u & k'_{22}(u) & = 0.
\end{align}
At the end of the day we want to obtain on-shell overlap formulas and we know that the on-shell Bethe vectors are highest weight states i.e.
\begin{equation}
 S^+ \left|\mathbf{u} \right\rangle = 0,
\end{equation}
therefore
\begin{equation}
 \bigl\langle\Psi\bigr|\mathbf{u}\bigr\rangle = 
 \bigl\langle\Psi'\bigr|\exp\left(\frac{c-\gamma}{a} S^+\right)\bigr|\mathbf{u}\bigr\rangle =
 \bigl\langle\Psi'\bigr|\mathbf{u}\bigr\rangle,
\end{equation}
where  the two-site state $\bigl\langle\Psi\bigr|$ is built from the $K$-matrix $\vec{K}'_{12}$. The
last equation shows that we can work with this rotated $K$-matrix for which
\begin{equation}
\kappa'(u) = \frac{b+\gamma u}{au}, \qquad \delta' = 0.
\end{equation}

\subsection{Recursion relation for the overlap}

Let us continue with the derivation of a recursion formula for the off-shell formula
\begin{equation}
\mathcal{S}_{N}(u_{1},\dots,u_{N})=\bigl\langle\Psi'\bigr|u_{1},\dots,u_{N}\bigr\rangle.
\end{equation}
We only need the following commutation relations \cite{Korepin-Book}
\begin{align}
D(u_{1})B(u_{2})\dots B(u_{N})= & \prod_{k=2}^{N}\frac{u_{1}-u_{k}+1}{u_{1}-u_{k}}B(u_{2})\dots B(u_{N})D(u_{1})-\nonumber \\
- & \sum_{l=2}^{N}\frac{1}{u_{1}-u_{l}}\prod_{k=2\neq l}^{N}\frac{u_{l}-u_{k}+1}{u_{l}-u_{k}}B(u_{1})\dots\widehat{B(u_{l})}\dots B(u_{N})D(u_{l}).
\end{align}

Let us start with the KT relation
\begin{equation}
\mathcal{S}_{N}(u_{1},\dots,u_{N})=\left\langle \Psi'\right|B(u_{1})B(u_{2})\dots B(u_{N})\left|0\right\rangle =\kappa'(u_{1}) \left\langle \Psi'\right|\left(D(-u_{1})-D(u_{1})\right)B(u_{2})\dots B(u_{N})\left|0\right\rangle.
\end{equation}
Substituting the commutation relations we can obtain the recursion
relation:
\begin{align}
\mathcal{S}_{N}(u_{1},\dots,u_{N})=\nonumber \\
\kappa'(u_{1})\biggl( & \prod_{k=2}^{N}\frac{u_{1}+u_{k}-1}{u_{1}+u_{k}}\lambda^{+}(u_{1})\mathcal{S}_{N-1}(u_{2},\dots,u_{N})-\nonumber \\
- & \prod_{k=2}^{N}\frac{u_{1}-u_{k}+1}{u_{1}-u_{k}}\lambda^{-}(u_{1})\mathcal{S}_{N-1}(u_{2},\dots,u_{N})+\nonumber \\
+ & \sum_{l=2}^{N}\frac{1}{u_{1}+u_{l}}\prod_{k=2\neq l}^{N}\frac{u_{l}-u_{k}+1}{u_{l}-u_{k}}\lambda^{-}(u_{l})\mathcal{S}_{N-1}(-u_{1},u_{2}\dots\widehat{u_{l}}\dots u_{N})+\nonumber \\
+ & \sum_{l=2}^{N}\frac{1}{u_{1}-u_{l}}\prod_{k=2\neq l}^{N}\frac{u_{l}-u_{k}+1}{u_{l}-u_{k}}\lambda^{-}(u_{l})\mathcal{S}_{N-1}(u_{1},u_{2}\dots\widehat{u_{l}}\dots u_{N})\biggr).\label{eq:rec}
\end{align}

We saw that the two magnon overlap is simplified as
\begin{align}
\mathcal{S}_{2}(u_{1},u_{2})=a^{\mathcal{L}}\kappa'(u_{1})\kappa'(u_{2})\biggl( & \frac{u_{1}+u_{2}-1}{u_{1}+u_{2}}\lambda^{+}(u_{1})\lambda^{+}(u_{2})-\frac{u_{1}-u_{2}+1}{u_{1}-u_{2}}\lambda^{-}(u_{1})\lambda^{+}(u_{2})-\nonumber \\
- & \frac{u_{1}-u_{2}-1}{u_{1}-u_{2}}\lambda^{+}(u_{1})\lambda^{-}(u_{2})+\frac{u_{1}+u_{2}+1}{u_{1}+u_{2}}\lambda^{-}(u_{1})\lambda^{-}(u_{2})\biggr).
\end{align}
Seeing this formula we may assume that the $N$ magnon overlap reads
as
\begin{equation}
\mathcal{S}_{N}(u_{1}\dots u_{N})=a^{\mathcal{L}}\left[\prod_{k=1}^{N}\kappa'(u_{k})\right]\left[\sum_{\left\{ \sigma_{1},\dots,\sigma_{N}\right\} =\left\{ \pm,\dots,\pm\right\} }\prod_{1\leq k<l\leq N}\frac{\sigma_{k}u_{k}+\sigma_{l}u_{l}-1}{\sigma_{k}u_{k}+\sigma_{l}u_{l}}\prod_{k=1}^{N}\sigma_{k}\lambda^{\sigma_{k}}(u_{k})\right].\label{eq:conj}
\end{equation}
It turns out that this assumption is true: the formula above can be proved using
the recursion relation (\ref{eq:rec}). We present the derivation 
in Appendix \ref{sec:Derivastion-of-off-shell}. 

The overlap formula (\ref{eq:conj}) can be also written in a determinant
form. After a simple rearrangement we obtain that
\begin{equation}
\mathcal{S}_{N}(u_{1}\dots u_{N})=
a^{\mathcal{L}}
\frac{\left[\prod_{k=1}^{N}\kappa'(u_{k})\right]}{\prod_{1\leq k<l\leq N}\left[u_{l}^{2}-u_{k}^{2}\right]}
\left[\sum_{\sigma_{1},\dots,\sigma_{N}}\prod_{k<l}\left[\left(u_{l}-\frac{\sigma_{l}}{2}\right)^{2}-\left(u_{k}-\frac{\sigma_{k}}{2}\right)^{2}\right]\prod_{k=1}^{N}\sigma_{k}\lambda^{\sigma_{k}}(u_{k})\right],
\end{equation}
which can be written as
\begin{equation}
\mathcal{S}_{N}(u_{1}\dots u_{N})=a^{\mathcal{L}}\left[\prod_{k=1}^{N}\kappa'(u_{k})\right]\frac{\det V_{N}^{+}(u_{1},\dots,u_{N})}{\prod_{1\leq k<l\leq N}\left[u_{l}^{2}-u_{k}^{2}\right]},\label{eq:offshellOverlap}
\end{equation}
where we defined a Vandermonde-like determinant 
\begin{equation}
  \label{vande1}
\left[V_{N}^{+}(u_{1},\dots,u_{N})\right]_{kl}=(u_{k}-1/2)^{2l-2}\lambda^{+}(u_{k})-(u_{k}+1/2)^{2l-2}\lambda^{-}(u_{k}).
\end{equation}
This result is reminiscent of the determinant formula found by Tsushiya
in \cite{tsushiya}, which describes the off-shell overlaps the
diagonal $K$-matrices in the spin-1/2 chain
\cite{sajat-neel}. However, the matrix elements are different already
in the spin-1/2 chain, with no direct connection to the formulas of
Tsushiya. Thus \eqref{eq:offshellOverlap} seems to be a completely new
result. We stress that it is valid not only in the spin-1/2 XXX chain,
but also in the higher spin integrable generalizations.

\subsection{On-shell overlaps}

In this subsection we review the connection between the integrability
condition and the pair structure of Bethe roots. Furthermore, we also derive the on-shell
overlap formulas, which are new in the higher spin cases.

We remind that the first derivation of exact on-shell overlaps of the
Heisenberg spin chains was presented in
\cite{Caux-Neel-overlap1}. In that work the on-shell limit of the off-shell formulas of
\cite{sajat-neel,sajat-karol} was taken, and the factorized structure
was found. The only work after \cite{Caux-Neel-overlap1,Caux-Neel-overlap2} which actually proved overlap
formulas was \cite{sajat-coba-proof}. In \cite{sajat-coba-proof} a new
method was introduced, which derived the on-shell overlaps based on
the analytic properties of the of-shell cases. Below we review this
method, and we use it to derive new on-shell overlaps.

Let us first discuss the implications of the integrability condition on Bethe roots.
Applying on-shell Bethe vectors on the integrability condition (\ref{eq:intCond})
we obtain that
\begin{equation}
\left\langle \Psi\right|t(u)\left|\mathbf{u}\right\rangle =\left\langle \Psi\right|\Pi t(u)\Pi\left|\mathbf{u}\right\rangle \quad\longrightarrow\quad\left(\Lambda(u)-\Lambda(-u)\right)\bigl\langle\Psi\bigr|\mathbf{u}\bigr\rangle=0,
\end{equation}
where $\Lambda(u)$ is the eigenvalue of the transfer matrix
\begin{equation}
t(u)\left|\mathbf{u}\right\rangle =\Lambda(u)\left|\mathbf{u}\right\rangle ,\qquad\Pi t(u)\Pi\left|\mathbf{u}\right\rangle =t(-u)\left|\mathbf{u}\right\rangle =\Lambda(-u)\left|\mathbf{u}\right\rangle .
\end{equation}
We just obtained that non-vanishing overlaps can be found only if 
\begin{equation}
\Lambda(u)=\Lambda(-u)\label{eq:Lamcond}
\end{equation}
The explicit form of the eigenvalue is
\begin{equation}
\Lambda(u)=\lambda^{+}(u)\frac{Q_{1}(u-1)}{Q_{1}(u)}+\lambda^{-}(u)\frac{Q_{1}(u+1)}{Q_{1}(u)}.
\end{equation}
Using the formula for $\Lambda(u)$ (\ref{eq:Lam}) we can convince
ourselves that the condition (\ref{eq:Lamcond}) is equivalent to
\begin{equation}
Q_{1}(-u)=(-1)^{N}Q_{1}(u)
\end{equation}
This implies that the set of (finite) Bethe roots is parity symmetric.
Generally the two possibilities are
\begin{equation}
\mathbf{u}=\left\{ u_{1},-u_{1},u_{2},-u_{2},\dots,u_{N/2},-u_{N/2}\right\} \quad\text{or}\quad\mathbf{u}=\left\{ u_{1},-u_{1},u_{2},-u_{2},\dots,u_{(N-1)/2},-u_{(N-1)/2},0\right\} .
\end{equation}

Let us now turn to the method of \cite{sajat-coba-proof} for the
derivation of on-shell overlaps. For simplicity we focus on the case
when all rapidities come in pairs. We also use the following notation for the two halves of the set
of Bethe roots:
\begin{align}
 \mathbf{u}^+_{N/2} &=\left\{ u_{1},u_{2},\dots,u_{N/2}\right\},\\
 \mathbf{u}^-_{N/2} &=\left\{ u_{N/2+1},u_{N/2+2},\dots,u_{N}\right\}.
\end{align}

The key observation of
\cite{sajat-coba-proof} was that the non-zero terms in the on-shell overlaps are obtained
from apparent poles of the off-shell overlaps associated with the pair
structure. The paper \cite{sajat-coba-proof} concentrated on
homogeneous spin chains. Let us introduce the notation
\begin{equation}
  w(u)=e^{iLp(u)}
\end{equation}
where $p(u)$ is the lattice momentum in the model under
consideration. In \cite{sajat-coba-proof} $w(u)$ was denoted as $a(u)$, but we chose to change the
notation to avoid confusion with the parameter $a$ of the $K$-matrix.

It was observed in \cite{sajat-coba-proof} that the
off-shell overlap has apparent poles of the type
\begin{equation}
  \frac{w(u_1)w(u_2)-1}{u_1+u_2}
\end{equation}
where we selected for simplicity a concrete pair
$(u_1,u_2)$. In the on-shell limit such factors will
acquire some finite value (depending on $L$) if
$u_1+u_2\to 0$. However, it was also shown in
\cite{sajat-coba-proof} that the regular terms all add up to zero, and
the finite on-shell overlap consists only of these pole
contributions. Thus it is necessary to understand the precise pole
structure of the off-shell overlaps.

Let us now turn to our new off-shell formulas. It is our goal to use
the arguments of \cite{sajat-coba-proof} to derive new on-shell
overlaps. In our case the $w$-variables are replaced by 
\begin{equation}
 w_k=\frac{\lambda^+(u_k)}{\lambda^-(u_k)},
\end{equation}
because we also treat the inhomogeneous cases. In the homogeneous case the original functions are reproduced. In the on-shell limit the Bethe Ansatz equations
\begin{equation}
 w_k \prod_{j=1}^N \frac{f(u_k-u_j)}{f(u_j-u_k)} = -1
\end{equation}
are satisfied where
\begin{equation}
 f(u) = \frac{u-1}{u}. 
\end{equation}

Returning to the representation \eqref{eq:conj} of the off-shell
overlaps we see the occurrence of the formal poles of the type $(u_1+u_2)^{-1}$. However, our
normalization is different from that 
of \cite{sajat-coba-proof}.
Let us therefore introduce the new normalization
\begin{equation}
 S_N(u_1,\dots,u_N) = \prod_{k=1}^N \frac{1}{\lambda^-(u_k)}  \mathcal{S}_N(u_1,\dots,u_N)
\end{equation}
with the exact result given by
\begin{equation}
 S_{N}(u_{1}\dots u_{N})=a^{\mathcal{L}}\left[\prod_{k=1}^{N}\kappa'(u_{k})\right]\left[\sum_{\left\{ \sigma_{1},\dots,\sigma_{N}\right\} =\left\{ \pm,\dots,\pm\right\} }\prod_{1\leq k<l\leq N}\frac{\sigma_{k}u_{k}+\sigma_{l}u_{l}-1}{\sigma_{k}u_{k}+\sigma_{l}u_{l}}\prod_{k=1}^{N}\sigma_{k}w_k^{\sigma_{k}+1/2} \right].
\end{equation}
We can see that this overlap function depends only on the $w_i$ parameters, and the separate
dependence on $\lambda^\pm(u_i)$ disappeared.

Let us regard the off-shell overlap as a function of the $u$-variables
and $w$-variables, which are treated as formally independent
variables.  
Taking the $u_{1}+u_{2}\to0$ limit we obtain the singular piece
\begin{multline}
S_{N}(u_{1}\dots u_{N})=
\frac{1-w_1w_2}{u_{1}+u_{2}}\kappa'(u_{1})\kappa'(-u_{1})\times \left[\prod_{k=3}^{N}\kappa(u_{k})\right]
\left[ \prod_{k=3}^N \frac{u_1-u_k-1}{u_1-u_k} \frac{u_1+u_k+1}{u_1+u_k}\right] \times\\
a^{\mathcal{L}} 
\left[\sum_{\left\{ \sigma_{3},\dots,\sigma_{N}\right\} }\prod_{3\leq k<l\leq N}\frac{\sigma_{k}u_{k}+\sigma_{l}u_{l}-1}{\sigma_{k}u_{k}+\sigma_{l}u_{l}}
 \prod_{k=3}^{N} \sigma_{k} \left(\frac{u_{k}+u_{1}-1}{u_{k}+u_{1}+1}\frac{u_{k}-u_{1}-1}{u_{k}-u_{1}+1}\right)^{\sigma_k+1/2}
  w^{\sigma_k+1/2}_k \right]\\
+\mathrm{reg.}
\end{multline}
Therefore the residue can be written as
\begin{multline}
  \label{prop1}
 S_{N}(u_{1}\dots u_{N}|w_1,\dots,w_N)=
 \frac{w_1w_2-1}{u_{1}+u_{2}} F(u_1) \times
 \left[ \prod_{k=3}^N f(u_1-u_k)f(-u_1-u_k)\right] \times\\
 S_{N-2}(u_{3}\dots u_{N}|\tilde{w}_3,\dots,\tilde{w}_N)+\mathrm{reg.}
\end{multline}
where $S_{N-2}$ contains the modified $w_k$ parameters
\begin{align}
 \tilde{w}_k & = \frac{f(u_{k}+u_{1})}{f(-u_{k}-u_{1})}\frac{f(u_{k}-u_{1})}{f(u_{1}-u_{k})} w_k.
\end{align}
Above $F(u)$ is a rational function which carries the dependence
on the initial state:
\begin{equation}
 F(u)= - \kappa'(u)\kappa'(-u) = - \frac{\gamma^2}{a^2} \frac{u^2-b^2/\gamma^2}{u^2}. 
\end{equation}
The structure of the residue in \eqref{prop1} agrees with Proposition 1 of \cite{sajat-coba-proof}.
Repeating the derivation of \cite{sajat-coba-proof} one can show that if the
above residue property holds, then the un-normalized on-shell overlap is equal to 
\begin{equation}
  S_N \to  a^{\mathcal{L}}\prod_{j=1}^{N/2}F(u_j^+)
\prod_{1\le j<k\le N/2}\bar f(u_j^+,u_k^+)
\times \det G^+_{N/2}. \label{eq:onshellov}
\end{equation}
where $G^{\pm}$ are the so-called Gaudin-like matrices of size
$\tfrac{N}{2}\times\tfrac{N}{2}$, with matrix elements
  \begin{equation}
    \label{eq:Gpm}
  G^\pm_{jk}=\left(\frac{\partial}{\partial u^+_j} \phi(u^+_k) \pm 
  \frac{\partial}{\partial u^+_j} \phi(u^-_k)\right) \Biggr|_{u^-_i=-u^+_i},
\end{equation}
with
\begin{equation}
 \phi(u) = \log \left[ \frac{\lambda^+(u)}{\lambda^-(u)} 
 \prod_{j=1}^N \frac{f(u-u_j)}{f(u_j-u)}\right]
\end{equation}
  and
\begin{equation}
  \bar f(\lambda,\mu)=f(\lambda-\mu)f(\lambda+\mu)f(-\lambda-\mu)f(-\lambda+\mu).
\end{equation} 

Therefore we just proved that the overlap function has the following on-shell limit
\begin{equation}
\langle \Psi |\mathbf{u}^+_{N/2} \rangle = a^{\mathcal{L}}\prod_{j=1}^{N/2}F(u_j^+)
\prod_{j=1}^{N/2} \lambda^+(u^+_j)\lambda^-(u^+_j) 
\prod_{1\le j<k\le N/2}\bar f(u_j^+,u_k^+)
\times \det G^+_{N/2}. \label{eq:onshell1}
\end{equation}

To calculate the normalized overlap we need the pairing between left and right eigenstates
\begin{equation}
\langle \mathbf{u} |\mathbf{u} \rangle = 
\prod_{j=1}^{N} \lambda^+(u_j)\lambda^-(u_j) 
\prod_{1\le j<k\le N} f(u_j-u_k)f(u_k-u_j)
\times \det G_{N}.
\end{equation}
where $G$ is the Gaudin matrix
\begin{equation}
  G_{jk}=\frac{\partial}{\partial u_j} \phi(u_k).
\end{equation}
For the pair structure the Gaudin determinant is factorized
\begin{equation}
 \det G_N = \det G^+_{N/2} \det G^-_{N/2}. 
\end{equation}
We can then obtain the normalized overlap as
\begin{equation}
\frac{\skalarszorzat{\Psi}{\lanp}\skalarszorzat{\lanp}{\Psi}}{\skalarszorzat{\lanp}{\lanp}}
= a^{2\mathcal{L}}
\prod_{j=1}^{N/2} \nu(u_j^+)
\frac{\det G^+_{N/2}}{\det G^-_{N/2}}, \label{eq:onshell2}
\end{equation}
where the one particle overlap function is
\begin{equation}
 \nu(u)=\frac{(F(u))^2}{f(2u)f(-2u)} = 
 \frac{\gamma^4}{a^4} \frac{(u^2-b^2/\gamma^2)^2}{u^2(u^2-1/4)}. 
\end{equation}
We observe the remarkable factorized form of the on-shell overlap. It is important that the formula
\eqref{eq:onshell2} is valid also in the integrable higher spin chains with $SU(2)$-symmetry. The
dependence on the spin representation is carried only through the Gaudin-like matrices, because their
entries depend on the functions $\lambda^\pm(u)$, which on the other hand depend on the spin, see
\eqref{lambdapm}. In contrast, the pair amplitude $\nu(u)$ is completely independent of the spin.
The algebraic reason behind this remarkable separation and factorization is that in
the higher spin cases the $K$-matrices are constructed using the fusion relation \eqref{eq:Kfusion},
and they are completely determined by the $K$-matrices of the defining representation.

An alternative interpretation of the factorized formula (including the higher spin cases) can be given
using the Quantum Transfer Matrix (QTM) method, see
\cite{sajat-integrable-quenches,sajat-mps,sajat-twisted-yangian}. We do not pursue that method here,
nevertheless let us give a few comments. In the QTM method the fused $K$-matrices
describe the fusion of boundary transfer matrices, which belong to the same commuting
hierarchy. It can then 
be argued using the methods of \cite{sajat-minden-overlaps}, that the pair amplitude $\nu(u)$ has to
be the same for all spins. The argument using the QTM method can be applied only in the infinite
volume limit, but our computations here reach the same conclusions, using rigorous computations in
arbitrary finite volume.

\subsection{Descendant states}

We close this section by calculating the overlaps for the descendant states
\begin{equation}
\bigl\langle\Psi\bigr|\mathbf{u},M\bigr\rangle.
\end{equation}

\subsection*{Overlaps with $\bigl\langle\Psi'\bigr|$}

Let us start with the two-site state $\bigl\langle\Psi'\bigr|$, which
is obtained from the rotated $K$-matrix \eqref{rotK} and for
which the off-shell formula is available. For an off-shell state we can continuously change the rapidities, which means that we can obtain the descendant overlaps from a simple limit
\begin{equation}
\bigl\langle\Psi'\bigr|u_{1}\dots,u_{N},M\bigr\rangle=\lim_{u_{N+1},\dots,u_{N+M}\to\infty}\prod_{i=1}^{M}\frac{1}{u_{N+i}^{2L-1}}S_{N+M}(u_{1},\dots,u_{N+M}).
\end{equation}
Let us start with the case $M=1$. Now
\begin{equation}
S_{N+1}(u_{1},\dots,u_{N+1})=a^{\mathcal{L}}\left[\prod_{k=1}^{N+1}\kappa'(u_{k})\right]\left[\sum_{\left\{
      \sigma_{1},\dots,\sigma_{N+1}\right\} =\left\{
      \pm,\dots,\pm\right\}
  }f_{N+1}(\sigma_{1}\dots,\sigma_{N+1})\right],
\label{eq:ovN+1}
\end{equation}
where
\begin{equation}
f_{N+1}(\sigma_{1}\dots,\sigma_{N+1})=\prod_{1\leq k<l\leq N+1}\frac{\sigma_{k}u_{k}+\sigma_{l}u_{l}-1}{\sigma_{k}u_{k}+\sigma_{l}u_{l}}\prod_{k=1}^{N+1}\sigma_{k}\lambda^{\sigma_{k}}(u_{k}).
\end{equation}
We can pair the terms as
\begin{multline}
f_{N+1}(\sigma_{1}\dots,\sigma_{N},+)+f_{N+1}(\sigma_{1}\dots,\sigma_{N},-)=\prod_{1\leq k<l\leq N}\frac{\sigma_{k}u_{k}+\sigma_{l}u_{l}-1}{\sigma_{k}u_{k}+\sigma_{l}u_{l}}\prod_{k=1}^{N}\sigma_{k}\lambda^{\sigma_{k}}(u_{k})\times\\
\left[\prod_{1\leq k\leq N}\left(1-\frac{1}{u_{N+1}+\sigma_{k}u_{k}}\right)\lambda^{+}(u_{N+1})-\prod_{k=1}^{N}\left(1+\frac{1}{u_{N+1}-\sigma_{k}u_{k}}\right)\lambda^{-}(u_{N+1})\right]=\\
f_{N}(\sigma_{1}\dots,\sigma_{N})u_{N+1}^{2L-1}\left[2\left(\mathcal{L}-N\right)+\mathcal{O}(u_{N+1}^{-1})\right],
\label{eq:asym}
\end{multline}
where we used the following asymptotic expansions
\begin{align}
u^{-2L}\lambda^{-}(u) & =1+\pm\mathcal{L}u^{-1}+\mathcal{O}(u^{-2}),\quad\\
\prod_{k=1}^{N}\left(1\mp\frac{1}{u\pm\sigma_{k}u_{k}}\right) & =1\mp Nu^{-1}+\mathcal{O}(u^{-2}).
\end{align}
Substituting (\ref{eq:asym}) to (\ref{eq:ovN+1}) we obtain that
\begin{equation}
\lim_{u_{N+1}\to\infty}\frac{1}{u_{N+1}^{2L-1}}S_{N+1}(u_{1},\dots,u_{N+1})=2\left(\mathcal{L}-N\right)\frac{\gamma}{a}S_{N}(u_{1},\dots,u_{N}). 
\end{equation}
Using this formula we get the following off-shell result
\begin{equation}
\bigl\langle\Psi'\bigr|u_{1}\dots,u_{N},M\bigr\rangle=\left(\frac{2\gamma}{a}\right)^{M}\prod_{i=0}^{M-1}\left(\mathcal{L}-N-i\right)S_{N}(u_{1},\dots,u_{N}). \label{eq:infLimit}
\end{equation}
Therefore the ratio of off-shell formulas can be written as
\begin{equation}
\frac{\bigl\langle\Psi'\bigr|\mathbf{u},M\bigr\rangle}{\bigl\langle\Psi'\bigr|\mathbf{u}\bigr\rangle}=\left(\frac{2\gamma}{a}\right)^{M}\prod_{i=0}^{M-1}\left(\mathcal{L}-N-i\right)=\left(\frac{2\gamma}{a}\right)^{M}\frac{\left(\mathcal{L}-N\right)!}{\left(\mathcal{L}-N-M\right)!}.
\end{equation}
This result holds also in the on-shell limit.

\subsection*{Overlaps with $\bigl\langle\Psi\bigr|$}

Let us continue with the general case. Since we have no off--shell
formula we have to do something different as in the previous case.
We can only use the on-shell formula
\begin{equation}
\bigl\langle\Psi\bigr|\mathbf{u}\bigr\rangle=a^{\mathcal{L}}\prod_{i=1}^{N/2}F(u^+_{i})\times\mathcal{G}_{N/2},
\end{equation}
where $\mathcal{G}_{N}$ is independent form the two-site state
\begin{equation}
\mathcal{G}_{N/2} = \prod_{j=1}^{N/2} \lambda^+(u^+_j)\lambda^-(u^+_j) 
\prod_{1\le j<k\le N/2}\bar f(u_j^+,u_k^+)
\times \det G^+_{N/2}.
\end{equation}
and
\begin{equation}
F(u)=-\kappa'(u)\kappa'(-u)=-\frac{\gamma^{2}}{a^{2}}\frac{u^{2}-b^{2}/\gamma^{2}}{u^{2}}=\frac{\gamma^{2}}{a^{2}}h(u),
\end{equation}
where
\begin{equation}
 h(u)=-\frac{u^{2}-b^{2}/\gamma^{2}}{u^{2}}.
\end{equation}
Using these notations the overlap can be written as
\begin{equation}
\bigl\langle\Psi\bigr|\mathbf{u}\bigr\rangle=a^{\mathcal{L}-N}\gamma^{N}\prod_{i=1}^{N/2}h(u^+_{i})\times\mathcal{G}_{N/2}. \label{eq:onshelllim}
\end{equation}
Let us use the following expression for the descendant states
\begin{equation}
\left|\mathbf{u},M\right\rangle :=\frac{d^{M}}{d\alpha^{M}}\exp\left(\alpha\Delta(S^{-})\right)\Biggr|_{\alpha=0}\left|\mathbf{u}\right\rangle .
\end{equation}
Applying it to the overlap
\begin{equation}
\bigl\langle\Psi\bigr|\mathbf{u},M\bigr\rangle=\frac{d^{M}}{d\alpha^{M}}\bigl\langle\Psi\bigr|\exp\left(\alpha\Delta(S^{-})\right)\bigr|\mathbf{u}\bigr\rangle\Biggr|_{\alpha=0}=\frac{d^{M}}{d\alpha^{M}}\bigl\langle\Psi_{\alpha}\bigr|\mathbf{u}\bigr\rangle\Biggr|_{\alpha=0}.
\end{equation}
The rotated two-site state can be built from the rotated $K$-matrix
\begin{equation}
K_{12}(u|\alpha)=K_{12}(u)\left[\exp\left(\alpha S^{-}\right)\otimes\exp\left(\alpha S^{-}\right)\right]=\left(\begin{array}{cc}
a_{\alpha}u & -b_{\alpha}+c_{\alpha}u\\
b_{\alpha}+c_{\alpha}u & d_{\alpha}u
\end{array}\right),
\end{equation}
where
\begin{align}
a_{\alpha} & =a+2c\alpha+d\alpha^{2} & b_{\alpha} & =b\\
c_{\alpha} & =c+d\alpha & d_{\alpha} & =d.
\end{align}
From this rotated $K$-matrix we can build a two-site state $\bigl\langle\Psi_{\alpha}\bigr|$.
We observe that 
\begin{equation}
\gamma_{\alpha}^{2}=c_{\alpha}^{2}-a_{\alpha}d_{\alpha}=c^{2}-ad=\gamma^{2},
\end{equation}
therefore $\gamma$ and the function $h(u)$ are invariant quantities
w.r.t rotations. The rotated overlap can be written as
\begin{equation}
\frac{d^{M}}{d\alpha^{M}}\bigl\langle\Psi_{\alpha}\bigr|\mathbf{u}\bigr\rangle\Biggr|_{\alpha=0}=\frac{d^{M}}{d\alpha^{M}}\left[a_{\alpha}^{\mathcal{L}-N}\right]_{\alpha=0}\gamma^{N}\prod_{i=1}^{N/2}h(u^+_{i})\times\mathcal{G}_{N/2}.
\end{equation}
Using the expansion
\begin{equation}
a_{\alpha}^{\mathcal{L}-N}=(a+2c\alpha+d\alpha^{2})^{\mathcal{L}-N}=
\sum_{M=0}^{\mathcal{L}-N} \alpha^M
\sum_{i=0}^{\left\lfloor M/2\right\rfloor }\binom{\mathcal{L}-N}{i}\binom{\mathcal{L}-N-i}{M-2i}d^{i}(2c)^{M-2i}a^{\mathcal{L}-N-M+i}
\end{equation}
we obtain
\begin{equation}
\frac{d^{M}}{d\alpha^{M}}\left[a_{\alpha}^{\mathcal{L}-N}\right]_{\alpha=0}=a^{\mathcal{L}-N}A(N,M),
\end{equation}
where
\begin{equation}
A(N,M)=M!\sum_{i=0}^{\left\lfloor M/2\right\rfloor }\binom{\mathcal{L}-N}{i}\binom{\mathcal{L}-N-i}{M-2i}d^{i}(2c)^{M-2i}a^{i-M}.
\end{equation}
Therefore the overlaps for descendant states are
\begin{equation}
\bigl\langle\Psi\bigr|\mathbf{u},M\bigr\rangle=A(N,M)a^{\mathcal{L}-N}\gamma^{N}\prod_{i=1}^{N/2}h(u^+_{i})\times\mathcal{G}_{N/2}.
\end{equation}
We can see that the overlaps with  the Bethe state and its descendant
states are the same up to a numerical prefactor:
\footnote{For a special matrix product state it was already observed that the overlaps with the Bethe and their descendant states are the same up to a  combinatorical prefactor \cite{deLeeuw:2017dkd}.}  
\begin{equation}
\frac{\bigl\langle\Psi\bigr|\mathbf{u},M\bigr\rangle}{\bigl\langle\Psi\bigr|\mathbf{u}\bigr\rangle}=A(N,M).
\end{equation}
This is consistent with the previous result, since
\begin{equation}
A(N,M)\Biggr|_{\delta=0}=M!\binom{\mathcal{L}-N}{M}(2c)^{M}a^{-M}=\left(\frac{2c}{a}\right)^{M}\frac{\left(\mathcal{L}-N\right)!}{\left(\mathcal{L}-N-M\right)!}.
\end{equation}

\section{Integrable final states for twisted spin chains}

So far the integrable initial states and the exact overlaps have not yet been
considered in spin chains with twisted boundary conditions. The main
reason for this is that the observed properties of the integrable
states, such as the ``pair structure'' for the overlaps seemed to be
tied to the periodic boundary conditions. Here we show that there is a
natural generalization to twisted cases, given that the twist is
compatible with the $K$-matrix. Furthermore, we also derive new
overlap formulas. The computations in this Section will
form the basis of the SoV treatment in Section \ref{sec:sov}.

\subsection{Twisted spin chains}

The transfer matrix for the twisted case can be written as
\begin{equation}
t(u)=\mathrm{Tr}_{0}\left[T_{0}(u)G_{0}\right],
\end{equation}
where $G$ is the twist matrix. For the simplicity we use a diagonal
twist
\begin{equation}
G=\left(\begin{array}{cc}
z_{1} & 0\\
0 & z_{2}
\end{array}\right).
\end{equation}
The off-shell Bethe vectors have the same form as the untwisted ones
\begin{equation}
\left|\mathbf{u}\right\rangle =B(u_{1})\dots B(u_{N})\left|0\right\rangle .
\end{equation}
The only difference is that now the eigenvalue of the transfer matrix is
\begin{equation}
\Lambda(u)=z_{1}\lambda^{+}(u)\frac{Q_{1}(u-1)}{Q_{1}(u)}+z_{2}\lambda^{-}(u)\frac{Q_{1}(u+1)}{Q_{1}(u)}.
\end{equation}
There is an alternative way to build the eigenstates
\begin{equation}
\left|\mathbf{v}\right\rangle =C(v_{1})\dots C(v_{2\mathcal{L}-N})\left|0'\right\rangle,
\end{equation}
where $\left|0'\right\rangle $ is the lowest weight reference state
\begin{equation}
\left|0'\right\rangle =\left|0,s_{1}\right\rangle ^{(s_{1})}\otimes\left|0,s_{1}\right\rangle ^{(s_{1})}\otimes\dots\otimes\left|0,s_{L}\right\rangle ^{(s_{L})}\otimes\left|0,s_{L}\right\rangle ^{(s_{L})}.
\end{equation}
The new vector $\left|\mathbf{v}\right\rangle $ is an eigenvector
of $t(u)$ with the same eigenvalue as $\left|\mathbf{u}\right\rangle $
when the $QQ$-relation is satisfied
\begin{equation}
z_{2}Q_{1}\left(u+\frac{1}{2}\right)Q_{2}\left(u-\frac{1}{2}\right)-z_{1}Q_{1}\left(u-\frac{1}{2}\right)Q_{2}\left(u+\frac{1}{2}\right)=(z_{2}-z_{1})Q_{12}\left(u\right),
\end{equation}
where
\begin{multline}
Q_{12}\left(u\right)=\prod_{k=1}^{L}\left(u^{2}-\left(\theta_{k}+\frac{s_{k}-1}{2}\right)^{2}\right)\left(u^{2}-\left(\theta_{k}+\frac{s_{k}-3}{2}\right)^{2}\right)\times\cdots\\
\dots\times\left(u^{2}-\left(\theta_{k}+\frac{-s_{k}+3}{2}\right)^{2}\right)\left(u^{2}-\left(\theta_{k}+\frac{-s_{k}+1}{2}\right)^{2}\right).
\end{multline}
In summary, the eigenvalues can be expressed in two formally different ways
\begin{align}
\Lambda(u) & =z_{1}\lambda^{+}(u)\frac{Q_{1}(u-1)}{Q_{1}(u)}+z_{2}\lambda^{-}(u)\frac{Q_{1}(u+1)}{Q_{1}(u)}\label{eq:LamQ1}\\
 & =z_{2}\lambda^{+}(u)\frac{Q_{2}(u-1)}{Q_{2}(u)}+z_{1}\lambda^{-}(u)\frac{Q_{2}(u+1)}{Q_{2}(u)},\label{eq:LamQ2}
\end{align}
which are however equivalent if the $QQ$-relation is satisfied.

Let us continue with the integrable states. In the untwisted case
the intuition of the integrability condition was that the action of
the conserved charges and the space reflected conserved charges on
the integrable state must be equal. However, this argument implicitly assumed
that the space reflection is a symmetry of the system, or equivalently
that the transfer matrix and the space reflected transfer matrix are commuting.
This is generally not true for twisted models.

On the other hand, we saw that we can also
use $t(-u)$ in the integrability definition instead of the space
reflected transfer matrix (these are equal for the untwisted case
but not in the twisted one) and the $t(-u)$ generate the same set
of conserved charges as $t(u)$. Therefore the natural generalization
of the integrability condition is
\begin{equation}
  \left\langle \Psi\right|t(u)=\left\langle \Psi\right|t(-u).
  \label{eq:intCond_twist}
\end{equation}
Importantly, $t(u)$ is now the twisted transfer matrix. The natural
question is: What is the implication of this definition on the Q-functions
or the  Bethe roots?

Applying the Bethe state $\left|\mathbf{u}\right\rangle $ on the integrability
condition we obtain that
\begin{equation}
\left(\Lambda(u)-\Lambda(-u)\right)\bigl\langle\Psi\bigr|\mathbf{u}\bigr\rangle=0,
\end{equation}
therefore the non-vanishing overlaps require that
\begin{equation}
  \Lambda(u)=\Lambda(-u).
  \label{eq:Lamcond-1}
\end{equation}
Using (\ref{eq:LamQ1})-(\ref{eq:LamQ2}) the left and right hand side can be written
as
\begin{align}
\Lambda(u) & =z_{1}\lambda^{+}(u)\frac{Q_{1}(u-1)}{Q_{1}(u)}+z_{2}\lambda^{-}(u)\frac{Q_{1}(u+1)}{Q_{1}(u)}\\
\Lambda(-u) & =z_{1}\lambda^{+}(u)\frac{Q_{2}(-u+1)}{Q_{2}(-u)}+z_{2}\lambda^{-}(u)\frac{Q_{2}(-u-1)}{Q_{2}(-u)}.
\end{align}
We can see that the condition (\ref{eq:Lamcond-1}) is equivalent
to
\begin{equation}
  \label{twisti1}
Q_{2}(-u)=(-1)^{N}Q_{1}(u).
\end{equation}
This condition obviously requires that there is a selection rule for
the number of magnons
\begin{equation}
N=\mathcal{L},
\end{equation}
and the set of $\mathbf{u}$ is the same as $\mathbf{v}$ but with
opposite signs
\begin{equation}
  \label{twisti2}
\left\{ u_{1},u_{2},\dots,u_{N}\right\} =\left\{ -v_{1},-v_{2},\dots,-v_{N}\right\} .
\end{equation}
This is the natural generalization of the pair structure to the twisted cases. In the original
un-twisted case the rapidities are paired with each other; here the rapidities from two different
representations of the same state are paired. Going back to the un-twisted case we get $-{\bf v}={\bf u}$ iff the pair
structure for ${\bf u}$ also holds, thus the different integrability conditions are indeed
equivalent. We stress that \eqref{twisti2} is very restrictive and it does not hold for an arbitrary
on-shell configuration.

Let us continue with the two site states which are solutions of the
integrability condition (\ref{eq:intCond_twist}). For the two-site
state 
\begin{equation}
\left\langle \Psi\right|=\vec{K}^{(s_{1})}(\theta_{1})\otimes\dots \otimes \vec{K}^{(s_{L})}(\theta_{L})
\end{equation}
we saw that it satisfies the KT-relation
\begin{equation}
\left\langle \Psi\right|\vec{K}_{12}(u)T_{1}(u)=\left\langle \Psi\right|\vec{K}_{12}(u)T_{2}^{\pi}(u).
\end{equation}
or equivalently
\begin{equation}
\left\langle \Psi\right|\sigma^{y}\mathbf{K}^{t}(u)T(u)=\left\langle \Psi\right|T(-u)\sigma^{y}\mathbf{K}^{t}(u).
\end{equation}
Using this equation the action of the twisted transfer matrix can
be written as
\begin{equation}
\left\langle \Psi\right|t(u)=\left\langle \Psi\right|\mathrm{Tr}\left[T(-u)\sigma^{y}\mathbf{K}^{t}(u)G\left(\sigma^{y}\mathbf{K}^{t}(u)\right)^{-1}\right],
\end{equation}
which means that the integrability condition is satisfied if
\begin{equation}
\sigma^{y}\mathbf{K}^{t}(u)G=G\sigma^{y}\mathbf{K}^{t}(u).
\end{equation}
This is an important compatibility condition between the twist and
the $K$-matrix. 

Solutions of the KYB equation also satisfying this condition can be
written as
\begin{align}
k_{11}(u) & =0 & k_{12}(u) & =-b+cu\\
k_{21}(u) & =b+cu & k_{22}(u) & =0.
\end{align}

Let us continue with the determination of the off-shell overlap formula.
Since the off-shell Bethe vectors are completely independent from the twist we
can apply the previous off-shell overlap formula (\ref{eq:offshellOverlap})
for the twisted chain. Due to the compatibility condition we have to take the limit $a\to0$.
In this limit the reflection factor $\kappa(u)$ goes to $\infty$. Taking
the proper limit we obtain that non-vanishing overlaps require that
\begin{equation}
N=\mathcal{L},
\end{equation}
and the final formula is 
\begin{equation}
\mathcal{S}_{\mathcal{L}}(u_{1}\dots
u_{\mathcal{L}})=\left[\prod_{k=1}^{\mathcal{L}}\tilde{\kappa}(u_{k})\right]\frac{\det
  V_{\mathcal{L}}^{+}(u_{1},\dots,u_{\mathcal{L}})}{\prod_{1\leq
    k<l\leq\mathcal{L}}\left[u_{l}^{2}-u_{k}^{2}\right]},
\label{eq:twistedOverlap}
\end{equation}
where the Vandermonde-like determinant is given by \eqref{vande1} and
\begin{equation}
\tilde{\kappa}(u)=\frac{b+cu}{u}.
\end{equation}
The overlap formula (\ref{eq:twistedOverlap}) is also well-defined
for on-shell states. Potential problems could only appear from the
poles, but in the twisted case the pair structure does generally not
hold and therefore generally $u_k^2-u_l^2\ne 0$.

We can also see that ratio of overlaps with different integrable states
only depends on the $\tilde \kappa$ functions, namely
\begin{equation}
\frac{\bigl\langle\Psi\bigr|\mathbf{u}\bigr\rangle}{\bigl\langle\Psi'\bigr|\mathbf{u}\bigr\rangle}=\prod_{k=1}^{\mathcal{L}}\frac{\tilde{\kappa}(u_{k})}{\tilde{\kappa}'(u_{k})}.
\end{equation}
Let $\bigl\langle\Psi_{0}\bigr|$ be the two-site state corresponding
to $b=1,c=0$. The state $\bigl\langle\Psi_{0}\bigr|$ is the generalization of the Dimer state to arbitrary spins. Using this state as reference state we can obtain the
general overlap as
\begin{equation}
\bigl\langle\Psi\bigr|\mathbf{u}\bigr\rangle=\prod_{k=1}^{\mathcal{L}}(cu_{k}+b)\bigl\langle\Psi_{0}\bigr|\mathbf{u}\bigr\rangle=(-c)^{\mathcal{L}}Q_{1}(-b/c)\bigl\langle\Psi_{0}\bigr|\mathbf{u}\bigr\rangle.
\end{equation}

\subsection{Untwisted limit}

Here we investigate the limit when the model is tuned back the original
untwisted case.

The twisted boundary condition restricts
the possible two-site states radically, because the components $a$ and $d$ are fixed to zero
by the compatibility with the twist. Therefore one might think that the
integrability definition is too restrictive, as we loose a number of
states which were integrable for the untwisted model. It might appear
that we have not found the proper generalization of the integrability
condition to the twisted case, and that the twisted case is not very useful to study the original
untwisted problem.

However, it turns
out that the twisted overlap formula \eqref{eq:twistedOverlap} already
contains all the information which is needed to reconstruct the most
general untwisted on-shell overlap formula. This happens because the
general two-site state can be obtained from the restricted one by a
rotation and the twisted overlap formula gives all the non-vanishing
untwisted overlaps for the Bethe states, as well as their descendants. 

Let us start with the behavior of the Bethe roots in the untwisted limit. It is 
common knowledge that as the twist is tuned back to zero, the 
Bethe roots go either to the solutions of the
untwisted Bethe equations, or they approach  infinity. We 
know that the non-vanishing overlap requires the pair structure of the
Bethe roots for periodic boundary conditions. This means that for
$z_{1},z_{2}\to1$ the Bethe roots have the following limit 
\begin{align}
u_{2a+1}+u_{2a} & \to0,\quad\text{ for }a=1,\dots,N/2,\\
u_{k} & \to\infty,\quad\text{for }k=N+1,\dots,\mathcal{L},
\end{align}
and the limit of the Bethe vectors reads as
\begin{equation}
\lim_{z_{1},z_{2}\to1}\prod_{i=1}^{\mathcal{L}-N}\frac{1}{u_{N+i}^{2L-1}}\mathcal{S}_{\mathcal{L}}(u_{1}\dots u_{\mathcal{L}})=\lim_{z_{1},z_{2}\to1}\prod_{i=1}^{\mathcal{L}-N}\frac{1}{u_{N+i}^{2L-1}}\bigl\langle\bar{\Psi}\bigr|u_{1},\dots,u_{\mathcal{L}}\bigr\rangle=\bigl\langle\bar{\Psi}\bigr|\mathbf{u},\mathcal{L}-N\bigr\rangle,
\label{eq:untwLimit1}
\end{equation}
where $\bigl\langle\bar{\Psi}\bigr|$ is the two-site state which
can be built from the $K$-matrix $\bar{K}(u)$ for which 
\begin{align}
\bar{k}_{11}(u) & =0 & \bar{k}_{12}(u) & =-b+\gamma u\\
\bar{k}_{21}(u) & =b+\gamma u & \bar{k}_{22}(u) & =0.
\end{align}
Taking the untwisted limit of the twisted formula we obtain that
\begin{multline}
\lim_{z_{1},z_{2}\to1}\prod_{i=1}^{\mathcal{L}-N}\frac{1}{u_{N+i}^{2L-1}}\mathcal{S}_{\mathcal{L}}(u_{1}\dots u_{\mathcal{L}})=\left(2\gamma\right)^{\mathcal{L}-N}\left(\mathcal{L}-N\right)!
\mathcal{S}_{N}(u_{1},\dots,u_{N})=\\
=2^{\mathcal{L}-N}\left(\mathcal{L}-N\right)!\gamma^{\mathcal{L}}\prod_{i=1}^{N/2}h(u_{i})\times\mathcal{G}_{N}(\mathbf{u}), \label{eq:untwLimit2}
\end{multline}
where we used the $\kappa'=\bar{\kappa}$ and $M=\mathcal{L}-N$ limit of the equations \eqref{eq:infLimit} and \eqref{eq:onshelllim} in the first and the second rows. 
From \eqref{eq:untwLimit1} and \eqref{eq:untwLimit2} we obtained the on-shell untwisted overlaps for the special state $\bigl\langle\bar{\Psi}\bigr|$
\begin{equation}
  \bigl\langle\bar{\Psi}\bigr|\mathbf{u},\mathcal{L}-N\bigr\rangle=2^{\mathcal{L}-N}\left(\mathcal{L}-N\right)!\gamma^{\mathcal{L}}\prod_{i=1}^{N/2}h(u_{i})\times\mathcal{G}_{N}(\mathbf{u}).
  \label{eq:untwist}
\end{equation}
The general K-matrix (\ref{eq:genK1}-\ref{eq:genK2}) can obtained from $\bar{K}$ as
\begin{equation}
K_{12}(u)=\bar{K}_{12}(u)\left(\exp(\alpha S^{-})\otimes\exp(\alpha S^{-})\right)\left(\exp(\beta S^{+})\otimes\exp(\beta S^{+})\right),
\end{equation}
where
\begin{equation}
\alpha=\frac{a}{2\gamma},\qquad\beta=\frac{c-\gamma}{a}.
\end{equation}
Therefore the general two-site state can be obtained by a rotation
as
\begin{equation}
\bigl\langle\Psi\bigr|=\bigl\langle\bar{\Psi}\bigr|\exp(\alpha\Delta(S^{-}))\exp(\beta\Delta(S^{+})),
\end{equation}
which means that the general on-shell untwisted overlap can be written
as
\begin{equation}
\bigl\langle\Psi\bigr|\mathbf{u}\bigr\rangle=\bigl\langle\bar{\Psi}\bigr|\exp(\alpha\Delta(S^{-}))\exp(\beta\Delta(S^{+}))\bigr|\mathbf{u}\bigr\rangle.
\end{equation}
The highest weight property of the Bethe state implies that $S^+$ acts
trivially, and we can use simply the expansion of the exponential of
$S^-$ to obtain
\begin{equation}
\bigl\langle\Psi\bigr|\mathbf{u}\bigr\rangle=\frac{\alpha^{\mathcal{L}-N}}{(\mathcal{L}-N)!}\bigl\langle\bar{\Psi}\bigr|\mathbf{u},\mathcal{L}-N\bigr\rangle,
\end{equation}
where it was used that the states with non-zero spin have vanishing overlap
with $\bigl\langle\bar{\Psi}\bigr|$. 

Substituting (\ref{eq:untwist})
we obtain that
\begin{equation}
\bigl\langle\Psi\bigr|\mathbf{u}\bigr\rangle=\frac{\alpha^{\mathcal{L}-N}}{(\mathcal{L}-N)!}2^{\mathcal{L}-N}\left(\mathcal{L}-N\right)!\gamma^{\mathcal{L}}\prod_{i=1}^{N/2}h(u_{i})\times\mathcal{G}_{N}(\mathbf{u})=a^{\mathcal{L}-N}\gamma^{N}\prod_{i=1}^{N/2}h(u_{i})\times\mathcal{G}_{N}(\mathbf{u}).
\end{equation}
which agrees with the result of the previous section. 

\section{Overlaps and SoV}

\label{sec:sov}

In this section we embed the integrable initial states into
the framework of Separation of Variables. We investigate the overlaps
between the SoV basis and the integrable two-site states.
We will see that the SoV techniques can be
used to derive the overlaps in question, but interestingly we obtain a
formula which is different from the previous result
\eqref{eq:twistedOverlap}.

The SoV approach was pioneered by Sklyanin \cite{sklyanin-selected}. The
idea of the SoV is to find a basis in which the
eigenvectors of the transfer matrix factorize into one particle
blocks. It turns out that the SoV basis can be found by diagonalizing
the $B$-operator. However, in the periodic case the $B$-operator is
nilpotent, therefore one usually introduces a twist. 
For a diagonal twist the $B$-operator is still
nilpotent, therefore a non-diagonal twist is required. In the SoV
construction it is advantageous if the spectrum of the $B$ operator is
non-degenerate and this is the situation in the XXX spin chain if the
inhomogeneities are in a generic position.

There exist a number of equivalent
realizations of the SoV construction, depending on the conventions for
the twists. For our purposes the most
convenient choice is the set of conventions of   \cite{Gromov-nested} which
uses a rotated version of the transfer matrix. Following
\cite{Gromov-nested} we call it the  ``good'' transfer
matrix. 

For simplicity we concentrate here on the case when
the quantum space is in the defining representation i.e. $s_{k}=1$
for $k=1,\dots,L$. At first we review of the construction of the
SoV basis following \cite{Gromov-nested}. This basis diagonalizes
the ``good'' $B$-operator. Let us define the ``good'' monodromy
matrix
\begin{equation}
  \label{Tgood}
\mathbb{T}(u)=U^{-1}T(u)GU
\end{equation}
and ``good'' B-operator
\begin{equation}
\mathbb{B}(u)=\mathbb{T}_{12}(u),\qquad\bar{\mathbb{B}}(u)=\frac{z}{z^{2}-1}\mathbb{B}(u),
\end{equation}
where we use the following conventions for the twist and $U$:
\begin{equation}
G=\left(\begin{array}{cc}
z & 0\\
0 & 1/z
\end{array}\right),\qquad U=\left(\begin{array}{cc}
\alpha & \alpha\\
0 & 1/\alpha
\end{array}\right).
\end{equation}
The operator $T(u)$ in \eqref{Tgood} is the original monodromy matrix
defined in \eqref{Tdef}.

The left/right eigenvectors of $\mathbb{B}$ form the left/right SoV
basis:
\begin{equation}
\bigl\langle h_{1}\dots h_{2L}\bigr|\bar{\mathbb{B}}(u)=\prod_{i=1}^{2L}\left(u-\xi_{i}+h_{i}\right)\bigl\langle h_{1}\dots h_{2L}\bigr|,\qquad\bar{\mathbb{B}}(u)\bigr|h_{1}\dots h_{2L}\bigr\rangle=\prod_{i=1}^{2L}\left(u-\xi_{i}+h_{i}\right)\bigr|h_{1}\dots h_{2L}\bigr\rangle,
\end{equation}
where $h_{i}=\pm\frac{1}{2}$. We choose the normalization as
\begin{equation}
\bigl\langle\mathbf{h}\bigr|0\bigr\rangle=1,\qquad\bigl\langle0'\bigr|\mathbf{h}\bigr\rangle=1.
\end{equation}
These vectors and co-vectors are orthogonal and their ``norm'' is
\cite{Gromov-nested}
\begin{equation}
\mu(\mathbf{h})=\frac{1}{\bigl\langle\mathbf{h}\bigr|\mathbf{h}\bigr\rangle}=\frac{1}{\alpha^{4L}(z^{2}-1)^{2L}}\prod_{i=1}^{2L}(2h_{i})\prod_{i<j}\frac{\xi_{i}-\xi_{j}+h_{i}-h_{j}}{\xi_{i}-\xi_{j}}.
\end{equation}
The off-shell Bethe vectors can be written as
\begin{equation}
\bigr|\mathbf{u}\bigr\rangle=\prod_{i=1}^{N}\bar{\mathbb{B}}(u_{i})\bigr|0\bigr\rangle,\qquad\bigl\langle\mathbf{v}\bigr|=\prod_{i=1}^{2L-N}\bigl\langle0'\bigr|\bar{\mathbb{B}}(v_{i}).
\end{equation}
The overlaps are
\begin{equation}
\bigl\langle\mathbf{h}\bigr|\mathbf{u}\bigr\rangle=\prod_{i=1}^{2L}Q_{1}(\xi_{i}-h_{i}),\qquad\bigl\langle\mathbf{v}\bigr|\mathbf{h}\bigr\rangle=\prod_{i=1}^{2L}Q_{2}(\xi_{i}-h_{i}).
\end{equation}
When the vector $\bigr|\mathbf{u}\bigr\rangle$ and co-vector $\bigl\langle\mathbf{v}\bigr|$
are right and left eigenvectors of the transfer matrix with the same
eigenvalue then the Q-functions satisfy the $QQ$-relation:
\begin{equation}
z^{2}Q_{1}(u-1/2)Q_{2}(u+1/2)-Q_{1}(u+1/2)Q_{2}(u-1/2)=(z^{2}-1)Q_{12}(u),
\end{equation}
where
\begin{equation}
Q_{12}(u)=\prod_{i=1}^{2L}(u-\xi_{i}).
\end{equation}
The overlap between off-shell Bethe states can be written as
\begin{multline}
\bigl\langle\mathbf{v}\bigr|\mathbf{u}\bigr\rangle=\sum_{\mathbf{h}}\mu(\mathbf{h})\bigl\langle\mathbf{v}\bigr|\mathbf{h}\bigr\rangle\bigl\langle\mathbf{h}\bigr|\mathbf{u}\bigr\rangle=
\frac{1}{\alpha^{4L}(z^{2}-1)^{2L}}\sum_{\mathbf{h}}\prod_{i<j}\frac{\xi_{i}-\xi_{j}+h_{i}-h_{j}}{\xi_{i}-\xi_{j}}
\times\\ \times \prod_{i=1}^{2L}2h_{i}Q_{1}(\xi_{i}-h_{i})Q_{2}(\xi_{i}-h_{i})=
\frac{1}{\alpha^{4L}(z^{2}-1)^{2L}}\prod_{i<j}\frac{1}{\xi_{j}-\xi_{i}}\det W_{2L},\label{eq:ovoffBethe}
\end{multline}
where the matrix $W$ is
\begin{equation}
\left[W_{2L}\right]_{ij}=(\xi_i+1/2)^{j-1}Q_{1}^{-}(\xi_{i})Q_{2}^{-}(\xi_{i})-
(\xi_{i}-1/2)^{j-1}Q_{1}^{+}(\xi_{i})Q_{2}^{+}(\xi_{i}),\quad\text{for }1\leq a,b\leq2L.
\end{equation}
We are interested in overlaps with integrable final states.  In
accordance with the previous Sections we require
that the inhomogeneities are in pairs:
\begin{equation}
\xi_{2i-1}=\theta_{i},\qquad\xi_{2i}=-\theta_{i}.
\end{equation}
Similarly as before, non-vanishing overlaps can be found if the $Q$-functions satisfy
the integrability condition
\begin{equation}
Q_{1}(u)=(-1)^{L}Q_{2}(-u).\label{eq:condQ}
\end{equation}
Let us use these restrictions to simplify the Vandermonde-like determinants.
The rows with odd and even indices can be written as
\begin{align}
\left[W_{2L}\right]_{2i-1,b}= & 
\left[(\theta_{i}+1/2)^{b-1}Q_{1}^{-}(\theta_{i})Q_{2}^{-}(\theta_{i})-
(\theta_{i}-1/2)^{b-1}Q_{1}^{+}(\theta_{i})Q_{2}^{+}(\theta_{i})\right]\\
\left[W_{2L}\right]_{2i,b}=(-1)^{b} & 
\left[(\theta_{i}+1/2)^{b-1}Q_{1}^{-}(\theta_{i})Q_{2}^{-}(\theta_{i})-
(\theta_{i}-1/2)^{b-1}Q_{1}^{+}(\theta_{i})Q_{2}^{+}(\theta_{i})\right].
\end{align}
Subtracting the rows $W_{2i-1,b}$ from $W_{2i,b}$ we obtain that
the determinant is factorized as
\begin{multline}
\det W_{2L}=\left|\begin{array}{ccccc}
W_{1,1} & W_{1,2} & W_{1,3} & W_{1,4} & \dots\\
-W_{1,1} & W_{1,2} & -W_{1,3} & W_{1,4} & \dots\\
W_{3,1} & W_{3,2} & W_{3,3} & W_{3,4} & \dots\\
-W_{3,1} & W_{3,2} & -W_{3,3} & W_{3,4} & \dots\\
\vdots & \vdots & \vdots & \vdots & \ddots
\end{array}\right|_{2L\times2L}=\left|\begin{array}{ccccc}
W_{1,1} & W_{1,2} & W_{1,3} & W_{1,4} & \dots\\
0 & 2W_{1,2} & 0 & 2W_{1,4} & \dots\\
W_{3,1} & W_{3,2} & W_{3,3} & W_{3,4} & \dots\\
0 & 2W_{3,2} & 0 & 2W_{3,4} & \dots\\
\vdots & \vdots & \vdots & \vdots & \ddots
\end{array}\right|_{2L\times2L}=\\
2^{L}\left|\begin{array}{ccc}
W_{1,1} & W_{3,3} & \ldots\\
W_{3,1} & W_{3,3} & \ldots\\
\vdots & \vdots & \ddots
\end{array}\right|_{L\times L}\times\left|\begin{array}{ccc}
W_{1,2} & W_{1,4} & \ldots\\
W_{3,2} & W_{3,4} & \ldots\\
\vdots & \vdots & \ddots
\end{array}\right|_{L\times L}.
\end{multline}
Thus we find
\begin{equation}
  \det W_{2L}=2^{L}\det W_{L}^{+}\det W_{L}^{-},
  \label{eq:faktorV}
\end{equation}
where
\begin{align}
\left[W_{L}^{+}\right]_{ab} & =
(\theta_{a}+1/2)^{2b-2}Q_{1}^{-}(\theta_{a})Q_{2}^{-}(\theta_{a})-
(\theta_{a}-1/2)^{2b-2}Q_{1}^{+}(\theta_{a})Q_{2}^{+}(\theta_{a}),\\
\left[W_{L}^{-}\right]_{ab} & =
(\theta_{a}+1/2)^{2b-1}Q_{1}^{-}(\theta_{a})Q_{2}^{-}(\theta_{a})-
(\theta_{a}-1/2)^{2b-1}Q_{1}^{+}(\theta_{a})Q_{2}^{+}(\theta_{a}),
\end{align}
for $1\leq a,b\leq L$.
Using the condition \eqref{eq:condQ} we can express the matrices using only the $Q_1$ function as
\begin{align}
\left[W_{L}^{+}\right]_{ab} & =(-1)^L\left[
(\theta_{a}+1/2)^{2b-2}Q_{1}^{-}(\theta_{a})Q_{1}^{+}(-\theta_{a})-
(\theta_{a}-1/2)^{2b-2}Q_{1}^{+}(\theta_{a})Q_{1}^{-}(-\theta_{a})\right],\\
\left[W_{L}^{-}\right]_{ab} & =(-1)^L\left[
(\theta_{a}+1/2)^{2b-1}Q_{1}^{-}(\theta_{a})Q_{1}^{+}(-\theta_{a})-
(\theta_{a}-1/2)^{2b-1}Q_{1}^{+}(\theta_{a})Q_{1}^{-}(-\theta_{a})\right].
\end{align}
Substituting this to the overlap between two off-shell
Bethe states (\ref{eq:ovoffBethe}) we obtain that
\begin{equation}
\bigl\langle\mathbf{v}\bigr|\mathbf{u}\bigr\rangle=\frac{(-1)^{L}}{\alpha^{4L}(z^{2}-1)^{2L}}\left[\prod_{a=1}^{L}\frac{1}{\theta_{a}}\right]\left[\prod_{a<b}\frac{1}{(\theta_{a}^{2}-\theta_{b}^{2})^{2}}\right]\det W_{L}^{+}\det W_{L}^{-}.
\end{equation}

\subsection{Overlap formulas}

Now we want to calculate the overlaps with the previously defined boundary
state. For $s_{i}=1$ this is the Dimer which can be written as
\begin{equation}
\mathbf{K}^{(1)}=\left(\begin{array}{cc}
0 & -1\\
1 & 0
\end{array}\right),
\end{equation}
and
\begin{equation}
\bigl\langle\Psi_{0}\bigr|= \vec{K}^{(1)} \otimes \dots \otimes \vec{K}^{(1)},\qquad \bigr|\Psi_{0}\bigr\rangle= \vec{K}^{(1)T} \otimes \dots \otimes \vec{K}^{(1)T}.
\end{equation}
For the non-vanishing overlaps with on-shell Bethe states the Q-function
have to satisfy the integrability condition (\ref{eq:condQ}). Using the $KT$-relation
(\ref{eq:KT}) we can also check that the operator $\mathbb{B}$ acts
on the Dimer state as
\begin{equation}
\bigl\langle\Psi_{0}\bigr|\mathbb{B}(u)=\bigl\langle\Psi_{0}\bigr|\mathbb{B}(-u),\qquad\mathbb{B}(u)\bigr|\Psi_{0}\bigr\rangle=\mathbb{B}(-u)\bigr|\Psi_{0}\bigr\rangle,
\end{equation}
which means the overlaps $\bigl\langle\Psi_{0}\bigr|\mathbf{h}\bigr\rangle$
and $\bigl\langle\mathbf{h}\bigr|\Psi_{0}\bigr\rangle$ are not zero
iff 
\begin{equation}
h_{2i-1}=-h_{2i},\qquad\text{for }i=1,\dots,L.\label{eq:selectionrule}
\end{equation}
We can see that the local degrees of freedom is halved, because for
each pair $(h_{2i-1},h_{2i})$ only two possibilities give non-zero
overlap. The vectors with $h_{2i-1}=h_{2i}$ lead to zero overlap
for all $i=1,\dots,L$. This halving of the degrees of freedom in this
particular case is the analogous to the ``pair structure'' on the
level of the Bethe roots.

It is now convenient to define a new notation for the independent
degrees of freedom as
\begin{equation}
\bigl\langle f_{1}\dots f_{L}\bigr|=\bigl\langle h_{1}h_{2}\dots h_{2L-1}h_{2L}\bigr|,\qquad\bigr|f_{1}\dots f_{L}\bigr\rangle=\bigr|h_{1}h_{2}\dots h_{2L-1}h_{2L}\bigr\rangle,
\end{equation}
where
\begin{align}
(h_{2i-1} & =+\frac{1}{2},h_{2i}=-\frac{1}{2})\longleftrightarrow f_{i}=+1\\
(h_{2i-1} & =-\frac{1}{2},h_{2i}=+\frac{1}{2})\longleftrightarrow f_{i}=-1.
\end{align}
Our goal is to calculate the overlap
\begin{equation}
\bigl\langle\Psi_{0}\bigr|\mathbf{u}\bigr\rangle=\sum_{\mathbf{h}}\mu(\mathbf{h})\bigl\langle\Psi_{0}\bigr|\mathbf{h}\bigr\rangle\bigl\langle\mathbf{h}\bigr|\mathbf{u}\bigr\rangle=\sum_{\mathbf{f}}\mu(\mathbf{f})\bigl\langle\Psi_{0}\bigr|\mathbf{f}\bigr\rangle\bigl\langle\mathbf{f}\bigr|\mathbf{u}\bigr\rangle.
\end{equation}
In the formula we already know the norm and the overlap $\bigl\langle\mathbf{f}\bigr|\mathbf{u}\bigr\rangle$:
\begin{align}
\mu(\mathbf{f}) & =\frac{1}{\alpha^{4L}(z^{2}-1)^{2L}}\prod_{a=1}^{L}\frac{2\theta_{a}+f_{a}}{2\theta_{a}}\prod_{a<b}\left(\frac{f_{a}\theta_{a}+f_{b}\theta_{b}+1}{f_{a}\theta_{a}+f_{b}\theta_{b}}\right)^{2},\label{eq:mu}\\
\bigl\langle\mathbf{f}\bigr|\mathbf{u}\bigr\rangle & =(-1)^{L}\prod_{a=1}^{L}Q_{1}(\theta_{a}-f_{a}/2)Q_{2}(\theta_{a}-f_{a}/2),\label{eq:fu}\\
\bigl\langle\mathbf{v}\bigr|\mathbf{f}\bigr\rangle & =(-1)^{L}\prod_{a=1}^{L}Q_{1}(\theta_{a}-f_{a}/2)Q_{2}(\theta_{a}-f_{a}/2).\label{eq:fv}
\end{align}
In the formulas above we used the integrability condition for the $Q$-functions \eqref{eq:condQ}.
In the Appendix \ref{sec:The-overlap-between} we derive that the
overlap $\bigl\langle\Psi_{0}\bigr|\mathbf{f}\bigr\rangle$ can be
written as
\begin{align}
\bigl\langle\Psi_{0}\bigr|\mathbf{f}\bigr\rangle & =
C_{r}\prod_{a=1}^{L}\frac{f_{a}2\theta_{a}}{2\theta_{a}+f_{a}}
\prod_{a<b}\frac{f_{a}\theta_{a}+f_{b}\theta_{b}}{f_{a}\theta_{a}+f_{b}\theta_{b}+1},\label{eq:ovr}\\
\bigl\langle\mathbf{f}\bigr|\Psi_{0}\bigr\rangle & =
C_{l}\prod_{a=1}^{L}\frac{f_{a}2\theta_{a}}{2\theta_{a}+f_{a}}
\prod_{a<b}\frac{f_{a}\theta_{a}+f_{b}\theta_{b}}{f_{a}\theta_{a}+f_{b}\theta_{b}+1},
\end{align}
where the product of the normalization factors is
\begin{equation}
C_{l}C_{r}=\alpha^{4L}(z^{2}-1)^{2L}\prod_{a=1}^{L}\frac{(\theta_{a}+1/2)(\theta_{a}-1/2)}{\theta_{a}^{2}}.
\end{equation}
Substituting (\ref{eq:mu}),(\ref{eq:fu}) and (\ref{eq:ovr}) to
the overlap formula we obtain that
\begin{multline}
\bigl\langle\Psi_{0}\bigr|\mathbf{u}\bigr\rangle=\frac{C_{r}(-1)^{L}}{\alpha^{4L}(z^{2}-1)^{2L}}\sum_{\mathbf{f}}\prod_{a<b}\frac{f_{a}\theta_{a}+f_{b}\theta_{b}+1}{f_{a}\theta_{a}+f_{b}\theta_{b}}\prod_{a=1}^{L}f_{a}Q_{1}(\theta_{a}-f_{a}/2)Q_{2}(\theta_{a}-f_{a}/2)=\\
\frac{C_{r}(-1)^{L}}{\alpha^{4L}(z^{2}-1)^{2L}}
\prod_{a<b}\frac{1}{\theta_{b}^{2}-\theta_{a}^{2}}
\sum_{\mathbf{f}}\prod_{a<b}\left(\left(\theta_{b}+f_{b}/2\right)^{2}-\left(\theta_{a}+f_{a}/2\right)^{2}\right)\prod_{a=1}^{L}f_{a}Q_{1}(\theta_{a}-f_{a}/2)Q_{2}(\theta_{a}-f_{a}/2).
\end{multline}
We can see that this formula can be written in a compact form using
the Vandermonde-like determinant $W^{+}_L$:
\begin{equation}
\bigl\langle\Psi_{0}\bigr|\mathbf{u}\bigr\rangle=\frac{C_{r}(-1)^{L}}{\alpha^{4L}(z^{2}-1)^{2L}}
\prod_{a<b}\frac{1}{\theta_{b}^{2}-\theta_{a}^{2}}\det W_L^{+}.
\end{equation}
Similarly we can derive the overlap for the co-vectors:
\begin{equation}
\bigl\langle\mathbf{v}\bigr|\Psi_{0}\bigr\rangle=\frac{C_{l}(-1)^{L}}{\alpha^{4L}(z^{2}-1)^{2L}}\prod_{a<b}\frac{1}{\theta_{b}^{2}-\theta_{a}^{2}}\det W_L^{+}.
\end{equation}
Using these formulas the normalized overlap reads as
\begin{equation}
\frac{\bigl\langle\Psi_{0}\bigr|\mathbf{u}\bigr\rangle\bigl\langle\mathbf{v}\bigr|\Psi_{0}\bigr\rangle}{\bigl\langle\mathbf{v}\bigr|\mathbf{u}\bigr\rangle}=(-1)^{L}\prod_{a=1}^{L}\frac{(\theta_{a}+q/2)(\theta_{a}-q/2)}{\theta_{a}}\frac{\det W_L^{+}}{\det W_L^{-}}.
\end{equation}
We used the ``generalized pair'' structure (or integrability
condition) $Q_{1}(u)=(-1)^{L}Q_{2}(-u)$ in the derivation of the
factorization of $\det W$ \eqref{eq:faktorV} and in
\eqref{eq:fu}-\eqref{eq:fv}, but we did not use the Bethe Ansatz or the
$QQ$-relations. This means that the above formula  holds for both \emph{
off-shell and on-shell states}. 

In the previous Section we saw that the on-shell overlap of general
two-site state can be obtained from the overlap of the reference state $\bra{\Psi_0}$
as
\begin{equation}
\frac{\bigl\langle\Psi\bigr|\mathbf{u}\bigr\rangle}{\bigl\langle\Psi_{0}\bigr|\mathbf{u}\bigr\rangle}=\frac{\bigl\langle\mathbf{v}\bigr|\Psi\bigr\rangle}{\bigl\langle\mathbf{v}\bigr|\Psi_{0}\bigr\rangle}=(-c)^{L}Q_{1}(-b/c)=(c)^{L}Q_{2}(b/c).
\end{equation}
therefore the general normalized overlap can be written as
\begin{equation}
  \label{sova}
\frac{\bigl\langle\Psi\bigr|\mathbf{u}\bigr\rangle\bigl\langle\mathbf{v}\bigr|\Psi\bigr\rangle}{\bigl\langle\mathbf{v}\bigr|\mathbf{u}\bigr\rangle}=(-c^{2})^{L}Q_{1}^{2}(b/c)\prod_{i=1}^{L}\frac{(\theta_{i}+q/2)(\theta_{i}-q/2)}{\theta_{i}}\frac{\det W_L^{+}}{\det W_L^{-}}.
\end{equation}
We emphasize that this formula is different from the previous result
\eqref{eq:twistedOverlap}, even though some of the ingredients might
look similar.  At this point the connection between these two results
is not clear, and it deserves further study.

Also, it would be important to compute the homogeneous limit of the final
formula \eqref{sova}. It is known that in SoV 
the computation of the homogeneous limit is often very challenging,
and usually it requires a separate study\footnote{An interesting computation was presented in
  Appendix R of \cite{yunfeng-structure-g}, where a known overlap formula was expressed in a form
  that is reminiscent of the SoV results, for example it involves only the
  $Q$-functions. Connections to our results are not yet clear.}. Thus we leave this task to a
future work.

\section{Conclusion}

In this paper we investigated integrable initial/final states and their
overlaps using the Algebraic Bethe Ansatz  and the Separation of
Variables method. We obtained a number of results, and we feel it is
worthwhile to give here a list of them:

Our main results are  {\it a)} the
$KT$-relation \eqref{eq:intCond} which leads to the recursion
relations for the overlap, {\it b)} the off-shell overlap formula
\eqref{vande1} valid for all integrable final states of the higher
spin models, {\it c)} the corresponding on-shell formula \eqref{eq:onshell1}-\eqref{eq:onshell2},
{\it d)}  a generalization of the integrability condition to twisted
spin chains, see \eqref{twisti1}-\eqref{twisti2}, {\it e)}
the corresponding overlap formula \eqref{eq:twistedOverlap}, valid on-shell and
off-shell, {\it f)} the SoV representation \eqref{eq:selectionrule} of the integrability
condition for the Dimer state, and {\it g)} the final overlap formula
\eqref{sova} within SoV, valid for arbitrary integrable states of the
spin-1/2 chain.

In our view the most important results are the KT relation, the
generalization of the integrability condition to the twisted chains
and also to
the SoV method. This opens the way to study nested systems using
Algebraic Bethe Ansatz (ABA) and
SoV.
Combining the $KT$-relation with the techniques of
\cite{Hutsalyuk:2016srn} used in ABA one could get recursive equations and sum
rules for the overlaps, similar to those obtained for the scalar
products of the Bethe vectors in the models with $\mathfrak{gl}(m|n)$
symmetry \cite{Hutsalyuk:2017tcx,Hutsalyuk:2017way}.  
Focusing on SoV, our present results are expressed using the $Q$-functions, which are the natural
building blocks to treat the exact operator spectrum of the
$\mathcal{N}=4$ SYM theory. The extension of our methods to nested
systems and long range spin chains is a promising direction.
Thus our present results constitute an important step towards the exact one-point functions in
the defect CFT.

One of the interesting open questions is to what extent the present
  methods are helpful for the overlaps in those cases, when the
  initial/final states are given by integrable MPS
  \cite{zarembo-neel-cite3,ADSMPS2,kristjansen-proofs,adscft-kristjansen-2017,charlotte-zarembo-beyond-scalars,sajat-mps}. 
We believe that the $KT$-relation could be established also in those
cases, but it would lead to considerably more complicated recursion
relations. At present it is not clear whether the methods could be
useful in practical computations.

\subsection*{Acknowledgments}

We acknowledge collaboration with Yunfeng Jiang and Deliang Zhong on
early stages of this project.
T.G. was supported by NKFIH grant K134946.

\appendix

\section{Derivation of the off-shell overlap formula (\ref{eq:conj})\label{sec:Derivastion-of-off-shell}}

In this section we prove the off-shell overlap formula (\ref{eq:conj}).
In the derivation the following expression will appear
\begin{multline}
E_{M}(v|u_{1},\dots,u_{M})=\prod_{k=1}^{M}\frac{v+u_{k}-1}{v+u_{k}}\frac{v-u_{k}}{v-u_{k}-1}+\\
\sum_{l=1}^{M}\left(\frac{\kappa(-v)}{\kappa(u_{l})}\frac{1}{v+u_{l}}-\frac{\kappa(v)}{\kappa(u_{l})}\frac{1}{v-u_{l}}\right)\frac{v-u_{l}}{v-u_{l}-1}\prod_{k=1\neq l}^{M}\frac{u_{l}-u_{k}+1}{u_{l}-u_{k}}\frac{u_{l}+u_{k}}{u_{l}+u_{k}+1},
\end{multline}
where $u_{k}\neq0$ and $u_{k}\neq u_{l}$ for $k\neq l$ and 
\begin{equation}
\kappa(u)=\frac{b+cu}{au}.
\end{equation}
At first, we prove that the following identity holds
\begin{equation}
E_{M}(v|u_{1},\dots,u_{M})=1.\label{eq:id}
\end{equation}
To prove this identity we have to check that the $E_{M}(v|u_{1},\dots,u_{M})$
as a meromorphic function of $v$ is equal to $1$. In the $v\to\infty$
limit, it is satisfied i.e.
\begin{equation}
\lim_{v\to\infty}E_{M}(v|u_{1},\dots,u_{M})=1.
\end{equation}
Therefore the identity (\ref{eq:id}) holds if the meromorphic function
$E_{M}(v)$ has no poles in the $v$-plane. The formal poles are $v=0,u_{n},-u_{n},u_{n}+1$.
It is easy to convince ourselves that residues vanish
\begin{align}
\underset{v=u_{n}}{\mathrm{Res}}E_{M}(v|u_{1},\dots,u_{M})= & 0\\
\underset{v=0}{\mathrm{Res}}E_{M}(v|u_{1},\dots,u_{M})= & \sum_{l=1}^{M}\left(\frac{-b/a}{\kappa(u_{l})}\frac{1}{u_{l}}+\frac{b/a}{\kappa(u_{l})}\frac{1}{u_{l}}\right)\frac{u_{l}}{u_{l}+1}\prod_{k=1\neq l}^{M}\frac{u_{l}-u_{k}+1}{u_{l}-u_{k}}\frac{u_{l}+u_{k}}{u_{l}+u_{k}+1}=0.\\
\underset{v=-u_{n}}{\mathrm{Res}}E_{M}(v|u_{1},\dots,u_{M})= & -\frac{2u_{n}}{2u_{n}+1}\prod_{k=1\neq n}^{M}\frac{-u_{n}+u_{k}-1}{-u_{n}+u_{k}}\frac{-u_{n}-u_{k}}{-u_{n}-u_{k}-1}+\nonumber \\
 & +\frac{2u_{n}}{2u_{n}+1}\prod_{k=1\neq n}^{M}\frac{u_{n}-u_{k}+1}{u_{n}-u_{k}}\frac{u_{n}+u_{k}}{u_{n}+u_{k}+1}=0\\
\underset{v=u_{n}+1}{\mathrm{Res}}E_{M}(v|u_{1},\dots,u_{M})= & \frac{2u_{n}}{2u_{n}+1}\prod_{k=1\neq n}^{M}\frac{u_{n}+u_{k}}{u_{n}+u_{k}+1}\frac{u_{n}-u_{k}+1}{u_{n}-u_{k}}+\nonumber \\
 & +\left(\frac{\kappa(-u_{n}-1)}{\kappa(u_{n})}\frac{1}{2u_{n}+1}-\frac{\kappa(u_{n}+1)}{\kappa(u_{n})}\right)\prod_{k=1\neq l}^{M}\frac{u_{n}-u_{k}+1}{u_{n}-u_{k}}\frac{u_{n}+u_{k}}{u_{n}+u_{k}+1}=0,
\end{align}
therefore the identity (\ref{eq:id}) is satisfied.

Now, we turn on to prove (\ref{eq:conj}). Here we use induction.
Let us assume that (\ref{eq:conj}) holds for $\mathcal{S}_{N-1}$
i.e.
\begin{equation}
\mathcal{S}_{N-1}(u_{1},\dots,u_{N-1})=\sum_{\left\{ \sigma_{1},\dots,\sigma_{N}\right\} =\left\{ \pm,\dots,\pm\right\} }f(\sigma_{1},\dots,\sigma_{N-1})\prod_{k=1}^{N-1}\sigma_{k}\lambda^{\sigma_{k}}(u_{k}),
\end{equation}
where
\begin{equation}
f(\sigma_{1},\dots,\sigma_{N})=a^{\mathcal{L}}\left[\prod_{k=1}^{N}\kappa(u_{k})\right]\left[\prod_{1\leq k<l\leq N}\frac{\sigma_{k}u_{k}+\sigma_{l}u_{l}-1}{\sigma_{k}u_{k}+\sigma_{l}u_{l}}\right].
\end{equation}
Using the formula of $\mathcal{S}_{N-1}$ and the recursion relation
(\ref{eq:rec}) we can obtain an expression for $\mathcal{S}_{N}$:
\begin{align}
\mathcal{S}_{N}(u_{1}\dots u_{N})=\kappa(u_{1})\Biggl[ & \prod_{k=2}^{N}\frac{u_{1}+u_{k}-1}{u_{1}+u_{k}}\lambda^{+}(u_{1})\mathcal{S}_{N-1}(u_{2}\dots u_{N})-\nonumber \\
- & \prod_{k=2}^{N}\frac{u_{1}-u_{k}+1}{u_{1}-u_{k}}\lambda^{-}(u_{1})\mathcal{S}_{N-1}(u_{2}\dots u_{N})+\label{eq:req-1}\\
+ & \sum_{l=2}^{N}\frac{1}{u_{1}+u_{l}}\prod_{k=2\neq l}^{N}\frac{u_{l}-u_{k}+1}{u_{l}-u_{k}}\lambda^{-}(u_{l})\mathcal{S}_{N-1}(-u_{1},u_{2}\dots\widehat{u_{l}}\dots u_{N})+\nonumber \\
+ & \sum_{l=2}^{N}\frac{1}{u_{1}-u_{l}}\prod_{k=2\neq l}^{N}\frac{u_{l}-u_{k}+1}{u_{l}-u_{k}}\lambda^{-}(u_{l})\mathcal{S}_{N-1}(u_{1},u_{2}\dots\widehat{u_{l}}\dots u_{N})\Biggr]=\nonumber \\
= & \sum_{\left\{ \sigma_{1},\dots,\sigma_{N}\right\} =\left\{ \pm,\dots,\pm\right\} }\tilde{f}(\sigma_{1},\dots,\sigma_{N-1})\prod_{k=1}^{N-1}\sigma_{k}\lambda^{\sigma_{k}}(u_{k}).\nonumber 
\end{align}
The proof is complete if we show that the coefficients $f$ and $\tilde{f}$
are the same. Let us collect the terms $\lambda^{+}(u_{1})\lambda^{-}(u_{2})\dots\lambda^{-}(u_{M})\lambda^{+}(u_{M+1})\dots\lambda^{+}(u_{N})$.
Dividing the coefficients $f$ and $\tilde{f}$ we obtain that
\begin{multline}
\frac{\tilde{f}(+,-,\dots,-,+,\dots,+)}{f(+,-,\dots,-,+,\dots,+)}=\prod_{k=2}^{M}\frac{u_{1}-u_{k}}{u_{1}-u_{k}-1}\frac{u_{1}+u_{k}-1}{u_{1}+u_{k}}+\\
+\sum_{l=2}^{M}\left(\frac{\kappa(-u_{1})}{\kappa(u_{l})}\frac{1}{u_{1}+u_{l}}-\frac{\kappa(u_{1})}{\kappa(u_{l})}\frac{1}{u_{1}-u_{l}}\right)\frac{u_{1}-u_{l}}{u_{1}-u_{l}-1}\prod_{k=2\neq l}^{M}\frac{u_{l}-u_{k}+1}{u_{l}-u_{k}}\frac{u_{l}+u_{k}}{u_{l}+u_{k}+1}=\\
=E_{M-1}(u_{1}|u_{2},\dots,u_{M})=1,
\end{multline}
where the last line we used the identity (\ref{eq:id}). Since the
recursion relation is symmetry under the permutations of $u_{2},\dots,u_{N}$
we just obtained that
\begin{equation}
\tilde{f}(+,\sigma_{2},\dots,\sigma_{N-1})=f(+,\sigma_{2},\dots,\sigma_{N-1}).
\end{equation}
Analogously we can handle the coefficients with $\sigma_{1}=+$:
\begin{multline}
\frac{\tilde{f}(-,-,\dots,-,+,\dots,+)}{f(-,-,\dots,-,+,\dots,+)}=\prod_{k=2}^{M}\frac{u_{1}+u_{k}}{u_{1}+u_{k}+1}\frac{u_{1}-u_{k}+1}{u_{1}-u_{k}}+\\
+\sum_{l=2}^{M}\left(\frac{\kappa(-u_{1})}{\kappa(u_{l})}\frac{1}{u_{1}+u_{l}}-\frac{\kappa(u_{1})}{\kappa(u_{l})}\frac{1}{u_{1}-u_{l}}\right)\frac{u_{1}+u_{l}}{u_{1}+u_{l}+1}\prod_{k=2\neq l}^{M}\frac{u_{l}-u_{k}+1}{u_{l}-u_{k}}\frac{u_{l}+u_{k}}{u_{l}+u_{k}+1}=\\
=E_{M-1}(-u_{1}|u_{2},\dots,u_{M})=1,
\end{multline}
and using the permutation symmetry we just obtained that
\begin{equation}
\tilde{f}(-,\sigma_{2},\dots,\sigma_{N-1})=f(-,\sigma_{2},\dots,\sigma_{N-1}).
\end{equation}
Thus we proved (\ref{eq:conj}).

\section{The overlap between the Dimer state and the SoV basis vectors\label{sec:The-overlap-between}}

The calculation of the overlap
$\bigl\langle\Psi_{0}\bigr|\mathbf{f}\bigr\rangle$ is
based on the integrability condition
\begin{equation}
\bigl\langle\Psi_{0}\bigr|t(u)\bigr|\mathbf{f}\bigr\rangle=\bigl\langle\Psi_{0}\bigr|t(-u)\bigr|\mathbf{f}\bigr\rangle.
\end{equation}
Therefore we need two to know how the transfer matrix act on the SoV
basis. The transfer matrix can be written as
\begin{equation}
t(u)=\mathbb{A}(u)+\mathbb{D}(u),
\end{equation}
where the operators $\mathbb{A}$ and $\mathbb{D}$ are the diagonal
elements of the ''good'' monodromy matrix 
\begin{equation}
\mathbb{A}(u)=\mathbb{T}_{11}(u),\qquad\mathbb{D}(u)=\mathbb{T}_{22}(u).
\end{equation}
Since the $R$-matrix is $\mathfrak{gl}(2)$ invariant ''good'' monodromy
is satisfy the RTT relation
\begin{equation}
R_{12}(u-v)\mathbb{T}_{1}(u)\mathbb{T}_{2}(v)=\mathbb{T}_{2}(v)\mathbb{T}_{1}(u)R_{12}(u-v),
\end{equation}
therefore
\begin{align}
(v-u+1)\mathbb{A}(v)\mathbb{B}(u) & =(v-u)\mathbb{B}(u)\mathbb{A}(v)+\mathbb{A}(u)\mathbb{B}(v)\\
(u-v+1)\mathbb{D}(v)\mathbb{B}(u) & =(u-v)\mathbb{B}(u)\mathbb{D}(v)+\mathbb{D}(u)\mathbb{B}(v).
\end{align}
Substituting $v=\xi_{i}-h_{i}$ we obtain that
\begin{align}
\mathbb{B}(u)\mathbb{A}(\xi_{i}-h_{i})\bigr|\mathbf{h}\bigr\rangle & =\frac{u-\xi_{i}+h_{i}-1}{u-\xi_{i}+h_{i}}\lambda_{\mathbf{h}}(u)\mathbb{A}(\xi_{i}-h_{i})\bigr|\mathbf{h}\bigr\rangle\\
\mathbb{B}(u)\mathbb{D}(\xi_{i}-h_{i})\bigr|\mathbf{h}\bigr\rangle & =\frac{u-\xi_{i}+h_{i}+1}{u-\xi_{i}+h_{i}}\lambda_{\mathbf{h}}(u)\mathbb{D}(\xi_{i}-h_{i})\bigr|\mathbf{h}\bigr\rangle.
\end{align}
Therefore the $\mathbb{A}(\xi_{i}-h_{i})$ and $\mathbb{D}(\xi_{i}-h_{i})$
are lowering and raising operator, respectively i.e.
\begin{align}
\mathbb{A}(\xi_{i}-1/2)\bigr|\dots,h_{i} & =+1/2,\dots\bigr\rangle=F_{i}^{(+\frac{1}{2})}\bigr|\dots,h_{i}=-1/2,\dots\bigr\rangle, & \mathbb{A}(\xi_{i}+1/2)\bigr|\dots,h_{i} & =-1/2,\dots\bigr\rangle=0,\\
\mathbb{D}(\xi_{i}+1/2)\bigr|\dots,h_{i} & =-1/2,\dots\bigr\rangle=F_{i}^{(-\frac{1}{2})}\bigr|\dots,h_{i}=+1/2,\dots\bigr\rangle, & \mathbb{D}(\xi_{i}-1/2)\bigr|\dots,h_{i} & =+1/2,\dots\bigr\rangle=0,
\end{align}
which means that at these special rapidities the transfer matrix acts on
the SoV basis as
\begin{equation}
t(\xi_{i}-h_{i})\bigr|\dots,h_{i},\dots\bigr\rangle=F_{i}^{(h_{i})}\bigr|\dots,-h_{i},\dots\bigr\rangle.
\end{equation}
We can easily fix the parameters $F_{i}^{(h_{i})}$. Let us consider the
following expression
\begin{equation}
  \bigl\langle0'\bigr|t(\xi_{i}-h_{i})\bigr|\dots,h_{i},\dots\bigr\rangle=F_{i}^{(h_{i})}.
  \label{eq:actionT}
\end{equation}
We also know the action of the transfer matrix on the pseudovacuum
$\bigl\langle0'\bigr|$
\begin{equation}
\bigl\langle0'\bigr|t(u)=zQ_{12}(u-1/2)+\frac{1}{z}Q_{12}(u+1/2),
\end{equation}
therefore
\begin{equation}
\bigl\langle0'\bigr|t(\xi_{i}-h_{i})\bigr|\dots,h_{i},\dots\bigr\rangle=z^{2h_{i}}Q_{12}(\xi_{i}-2h_{i}),
\end{equation}
which means
\begin{equation}
F_{i}^{(h_{i})}=z^{2h_{i}}Q_{12}(\xi_{i}-2h_{i}).
\end{equation}
Since the transfer matrix is a matrix valued polynomial with degree
$2L$ we can use the following interpolation
\begin{equation}
t(u)=t_{\infty}\prod_{i=1}^{2L}(u-\xi_{i}-h_{i})+\sum_{j=1}^{2L}t(\xi_{j}-h_{j})\prod_{i=1\neq j}^{2L}\frac{u-\xi_{i}+h_{i}}{\xi_{j}-h_{j}-\xi_{i}+h_{i}},
\end{equation}
where $t_{\infty}$ is proportional to the identity. Using this interpolation
formula and (\ref{eq:actionT}) we obtain that
\begin{equation}
t(u)\bigr|\mathbf{h}\bigr\rangle=
t_{\infty}\bigr|\mathbf{h}\bigr\rangle\prod_{i=1}^{2L}(u-\xi_{i}+h_{i})+
\sum_{j=1}^{2L}F_{i}^{(h_{i})}\bigr|\dots,-h_{i},\dots\bigr\rangle\prod_{i=1\neq j}^{2L}\frac{u-\xi_{i}+h_{i}}{\xi_{j}-h_{j}-\xi_{i}+h_{i}}.
\end{equation}
Substituting this into the integrability condition
\begin{equation}
\bigl\langle\Psi_{0}\bigr|t(\theta_{1}+1/2)\bigr|-\frac{1}{2},-\frac{1}{2},\dots\bigr\rangle=\bigl\langle\Psi_{0}\bigr|t(-\theta_{1}-1/2)\bigr|-\frac{1}{2},-\frac{1}{2},\dots\bigr\rangle
\end{equation}
 we obtain that
\begin{multline}
F_{1}^{(-\frac{1}{2})}\bigl\langle\Psi_{0}\bigr|+\frac{1}{2},-\frac{1}{2},\dots\bigr\rangle=-\frac{1}{2\theta_{1}}F_{1}^{(-\frac{1}{2})}\bigl\langle\Psi_{0}\bigr|+\frac{1}{2},-\frac{1}{2},\dots\bigr\rangle\\
+\frac{2\theta_{i}+1}{2\theta_{1}}F_{2}^{(-\frac{1}{2})}\bigl\langle\Psi_{0}\bigr|-\frac{1}{2},+\frac{1}{2},\dots\bigr\rangle\prod_{i=3}^{2L}\frac{\theta_{1}+\frac{1}{2}+\xi_{i}-h_{i}}{\theta_{1}-\frac{1}{2}+\xi_{i}-h_{i}}.
\end{multline}
Here we used the selection rules $h_{2b-1}=-h_{2b}$ for the non-vanishing
overlaps $\bigl\langle\Psi_{0}\bigr|\mathbf{h}\bigr\rangle$, see
(\ref{eq:selectionrule}). We can see that the quotient of overlaps
can be written as
\begin{multline}
\frac{\bigl\langle\Psi_{0}\bigr|+\frac{1}{2},-\frac{1}{2},\dots\bigr\rangle}{\bigl\langle\Psi_{0}\bigr|-\frac{1}{2},+\frac{1}{2},\dots\bigr\rangle}=\frac{F_{2}^{(-\frac{1}{2})}}{F_{1}^{(-\frac{1}{2})}}\prod_{i=3}^{2L}\frac{\theta_{1}+\frac{1}{2}+\xi_{i}-h_{i}}{\theta_{1}-\frac{1}{2}+\xi_{i}-h_{i}}=\\
-\frac{2\theta_{i}-1}{2\theta_{i}+1}\prod_{b=2}^{L}\frac{\theta_{1}+\frac{1}{2}+\theta_{b}-h_{2b-1}}{\theta_{1}+\frac{1}{2}+\theta_{b}+\frac{1}{2}}\frac{\theta_{1}-\frac{1}{2}-\theta_{b}-\frac{1}{2}}{\theta_{1}-\frac{1}{2}-\theta_{b}+h_{2b-1}}\frac{\theta_{1}+\frac{1}{2}-\theta_{b}+h_{2b-1}}{\theta_{1}+\frac{1}{2}-\theta_{b}+\frac{1}{2}}\frac{\theta_{1}-\frac{1}{2}+\theta_{b}-\frac{1}{2}}{\theta_{1}-\frac{1}{2}+\theta_{b}-h_{2b-1}}.
\end{multline}
Using the notation $\bigr|f_{1}\dots,f_{L}\bigr\rangle$ this equation
is simplified as
\begin{equation}
\frac{\bigl\langle\Psi_{0}\bigr|+1,f_{2},\dots,f_{L}\bigr\rangle}{\bigl\langle\Psi_{0}\bigr|-1,f_{2},\dots,f_{L}\bigr\rangle}=-\frac{2\theta_{i}-1}{2\theta_{i}+1}\frac{\theta_{1}+f_{b}\theta_{b}}{\theta_{1}+f_{b}\theta_{b}+1}\frac{-\theta_{1}+f_{b}\theta_{b}+1}{-\theta_{1}+f_{b}\theta_{b}},
\end{equation}
therefore the overlap between the Dimer state and the SoV basis basis
vectors can be written as
\begin{equation}
  \bigl\langle\Psi_{0}\bigr|\mathbf{f}\bigr\rangle=C_{r}\prod_{a=1}^{L}\frac{2\theta_{a}}{2f_{a}\theta_{a}+1}\prod_{a<b}\frac{f_{a}\theta_{a}+f_{b}\theta_{b}}{f_{a}\theta_{a}+f_{b}\theta_{b}+1},
  \label{eq:ovsovr}
\end{equation}
where $C_{r}$ is independent on the variables $f_{a}$.

Analogously way we can calculate the overlap with the co-vector basis
$\bigl\langle\mathbf{f}\bigr|\Psi_{0}\bigr\rangle$. The only difference
is the action of the transfer matrix on
$\bigl\langle\mathbf{h}\bigr|$, which is given by
\begin{equation}
\bigl\langle\dots,h_{i},\dots\bigr|t(\xi_{i}-h_{i})=G_{i}^{(h_{i})}\bigl\langle\dots,-h_{i},\dots\bigr|,
\end{equation}
where
\begin{equation}
G_{i}^{(h_{i})}=z^{-2h_{i}}Q_{12}(\xi_{i}-2h_{i}).
\end{equation}
Repeating the same calculation as above we obtain that
\begin{equation}
\frac{\bigl\langle+\frac{1}{2},-\frac{1}{2},\dots\bigr|\Psi_{0}\bigr\rangle}{\bigl\langle-\frac{1}{2},+\frac{1}{2},\dots\bigr|\Psi_{0}\bigr\rangle}=\frac{G_{2}^{(-\frac{1}{2})}}{G_{1}^{(-\frac{1}{2})}}\prod_{i=3}^{2L}\frac{\theta_{1}+\frac{1}{2}+\xi_{i}-h_{i}}{\theta_{1}-\frac{1}{2}+\xi_{i}-h_{i}}=\frac{\bigl\langle\Psi_{0}\bigr|+\frac{1}{2},-\frac{1}{2},\dots\bigr\rangle}{\bigl\langle\Psi_{0}\bigr|-\frac{1}{2},+\frac{1}{2},\dots\bigr\rangle},
\end{equation}
therefore
\begin{equation}
\bigl\langle\mathbf{f}\bigr|\Psi_{0}\bigr\rangle=C_{l}\prod_{a=1}^{L}\frac{2\theta_{a}}{2f_{a}\theta_{a}+1}\prod_{a<b}\frac{f_{a}\theta_{a}+f_{b}\theta_{b}}{f_{a}\theta_{a}+f_{b}\theta_{b}+1},\label{eq:ovsovl}
\end{equation}
where $C_{l}$ is independent of $f_{a}$-s. We can fix the product
$C_{l}C_{r}$ by calculating the norm $\bigl\langle\Psi_{0}\bigr|\Psi_{0}\bigr\rangle=2^{L}$
and inserting the full system
\begin{equation}
\bigl\langle\Psi_{0}\bigr|\Psi_{0}\bigr\rangle=\sum_{\mathbf{h}}\mu(\mathbf{h})\bigl\langle\Psi_{0}\bigr|\mathbf{h}\bigr\rangle\bigl\langle\mathbf{h}\bigr|\Psi_{0}\bigr\rangle=\sum_{\mathbf{f}}\mu(\mathbf{f})\bigl\langle\Psi_{0}\bigr|\mathbf{f}\bigr\rangle\bigl\langle\mathbf{f}\bigr|\Psi_{0}\bigr\rangle.
\end{equation}
Substituting (\ref{eq:mu}), (\ref{eq:ovsovr}) and (\ref{eq:ovsovl})
we obtain that
\begin{multline}
\bigl\langle\Psi_{0}\bigr|\Psi_{0}\bigr\rangle=\frac{C_{l}C_{r}}{\alpha^{4L}(z^{2}-1)^{2L}}\sum_{\mathbf{f}}\prod_{a=1}^{L}\frac{2\theta_{a}}{2\theta_{a}+f_{a}}=\frac{C_{l}C_{r}}{\alpha^{4L}(z^{2}-1)^{2L}}\prod_{a=1}^{L}\frac{2\theta_{a}}{4\theta_{a}^{2}-1}\sum_{\mathbf{f}}\prod_{a=1}^{L}(2\theta_{a}-f_{a})=\\
2^{L}\frac{C_{l}C_{r}}{\alpha^{4L}(z^{2}-1)^{2L}}\prod_{a=1}^{L}\frac{4\theta_{a}^{2}}{4\theta_{a}^{2}-1}.
\end{multline}
Therefore 
\begin{equation}
C_{l}C_{r}=\alpha^{4L}(z^{2}-1)^{2L}\prod_{a=1}^{L}\frac{(\theta_{a}+1/2)(\theta_{a}-1/2)}{\theta_{a}^{2}}.
\end{equation}

% \bibliographystyle{elsarticle-num}
%\bibliographystyle{utphys}
%\bibliography{pozsi-ov,overlap}

\providecommand{\href}[2]{#2}\begingroup\raggedright\endgroup

\end{document}